\documentclass[a4paper]{article}

\usepackage{amsthm,amsmath,amssymb,amsfonts}

\def \ring {{\cal R}}

\def \ord {\, \mbox{ord}}

\def\bbbc{{\mathbb C}}
\def\bbbz{{\mathbb Z}}

\def\cH{{\cal H}}
\def\cK{{\cal K}}
\def\cJ{{\cal J}}

\def\tr{\mbox{tr}}

\newtheorem{Def}{Definition}
\newtheorem{The}{Theorem}
\newtheorem{Pro}{Proposition}

\begin{document}
\bibliographystyle{alpha}
\title{On classification of integrable non-evolutionary equations}
\author{Alexander V. Mikhailov$^{+}\footnote{
On leave, Landau Institute for Theoretical Physics, Moscow,
Russia}$, Vladimir S. Novikov$^{\$ *}$\footnote{The work on this
paper was partialy carried out in IMSAS, University of Kent, while
the author held a Royal Society NATO/Chevening Postdoctoral
Fellowship}
and Jing Ping Wang$ ^\& $\\
$+$ Applied Mathematics Department, University of Leeds, UK\\
$\&$ Institute of Mathematics and Statistics, University of Kent,
UK\\ $\$ $ FEW, Vrije Universiteit Amsterdam, The Netherlands}

\maketitle

\begin{abstract}
We study partial differential equations of second order (in time)
that possess a hierarchy of infinitely many higher symmetries.
The famous Boussinesq equation is a member of this class after
the extension of the differential polynomial ring.
We develop the perturbative symmetry approach in symbolic representation.
Applying it, we classify the integrable equations
of 4th and 6th order (in the space derivative) equations,
as well as we have found three new 10th order integrable equations.
To prove the integrability we provide
the corresponding bi-Hamiltonian structures and recursion operators.
\end{abstract}

\section{Introduction}

The variety of integrable evolutionary equations, i.e. equations explicitly resolved with
respect to the first time derivative
\[ u_t=F(u,u_x,u_{xx},...,u_{xx...x})\, , \]
have been extensively studied and most comprehensive results of
classification of integrable equations have been obtained for this
class of equations (see for example \cite{mss,sw}).

In this paper we  study necessary conditions for the integrability of non-evolutionary
equations of the form
\begin{equation}\label{eq1}
u_{tt}=\alpha \partial_x ^p u+\beta \partial_x ^q
u_{t}+K(u,u_x,u_{xx},...,\partial^{p-1}_x u,u_t,u_{t\, x},u_{t\,
xx},...,\partial^{q-1}_x u_{t})\, , \qquad p>q\, ,\quad
\alpha,\beta\in\bbbc
\end{equation}
and analyse equations satisfying these conditions.
We call an equation to be integrable if
it does possess an infinite hierarchy of higher symmetries. The
famous Boussinesq equation \cite{zakharov}
\begin{equation}\label{bouss1}
u_{tt}=u_{xxxx}+(u^2)_{xx}
\end{equation}
belongs to this class (after the extension of the differential polynomial ring).
Recently all integrable equations of the form
\begin{equation}\label{shabclass}
u_{tt}=u_{xxx}+F(u,u_x,u_{xx},u_t,u_{tx})
\end{equation}
have been classified and comprehensively studied in \cite{hss}.

Equations of the sixth order ($p=6$) of the form (\ref{eq1}) have
been studied in connection with reductions of the Sato hierarchies
corresponding to KP, BKP and CKP equations \cite{satoKP,satoBKP}.
Sixth order equations and their Lax representations can also be
derived from the systems studied by Drinfeld and Sokolov
\cite{DS1,DS2}.

The aim of our paper is to look in depth in the structure of hidden higher symmetries,
to develop a method that would allow to solve the classification problem for integrable equations of relatively
low order. It is a natural development of the Perturbative Symmetry approach \cite{mn}
based on the standard symmetry approach \cite{mss} and symbolic method developed in \cite{jp}.
In the spirit of perturbation theory we define approximate infinitesimal symmetries and approximate recursion operators.
Equations that have approximate recursion operator possess an infinite hierarchy of approximate symmetries and it is
natural to call such equations approximately integrable. A similar idea of approximate integrability, based on approximate
symmetries, conservation laws and recursion operators was discussed in
the context of obstacles to  asymptotic integrability \cite{kodmik0,kodmik}. In this paper we formulate explicit
and easily verifiable criteria for approximate integrability. The cases of odd and even order $p$ equations
are essentially different. The general theory presented in Section 2 is suitable for the both cases, but in
this paper we apply the theory to even order equations.
The results of the odd order equations will be published somewhere else.

In Section 3 we show that the method proposed in this paper does
reproduce all known integrable cases of equation (\ref{eq1}) (for
$p=4,6$) as well as enable us to find three new 10th order
equations. We  present recursion, Hamiltonian, symplectic
operators and the corresponding Lax representations for integrable
equations of 6th and 10th order.

\section{Approximate symmetries and recursion operators}

A non-evolutionary equation
\begin{equation}\label{eqnon}
u_{tt}=K(u,u_x,u_{xx},...,\partial^n_x u,u_t,u_{t\, x},u_{t\, xx},...,\partial^m_x u_{t})\, ,
\end{equation}
can always be replaced by a system of two evolutionary equations
\begin{eqnarray}
\left\{\begin{array}{l}\label{evsys0}
u_t=v,\\
v_t=K(u,u_x,u_{xx},...,\partial^n_x u,v,v_{x},v_{xx},...,\partial ^m_x v).
\end{array}\right .
\end{eqnarray}

Non-evolution equation (\ref{eqnon}) may have other representations in the form of
an evolutionary system of equations. If $K=D_x (G)$,
where $D_x$ is a total $x$-derivative, then the system of evolutionary
equations
\begin{equation}\label{sys2}
u_t=v_x\, ,\qquad
v_t=G\, .
\end{equation}
also represents (\ref{eqnon}).

For example, eliminating variable $v$ from the following equations

Case 1:
\begin{equation}\label{sysbouss1}
u_t=v\, ,\qquad
v_t=u_{xxxx}+(u^2)_{xx}\, ,
\end{equation}

Case 2:
\begin{equation}\label{sysbouss2}
u_t=v_x\, ,\qquad
v_t=u_{xxx}+(u^2)_{x}\, ,
\end{equation}

Case 3:
\begin{equation}\label{sysbouss3}
u_t=v_{xx}\, ,\qquad
v_t=u_{xx}+u^2\, ,
\end{equation}
we receive the same Boussinesq equation (\ref{bouss1}) on variable $u$ (variable $v$
has different sense and scaling dimension for these three cases).

There is a subtle, but important difference between the above representations
of the Boussinesq equation (\ref{bouss1}). It is easy to verify that
the system (\ref{sysbouss1}) has only a finite number of local infinitesimal
symmetries (i.e. symmetries whose generators can be expressed in terms of $u,v$ and a
finite number of their derivatives) while equations (\ref{sysbouss2}) and (\ref{sysbouss3}) have
infinite hierarchies of local symmetries. The reason is simple - infinitesimal
symmetries of equation (\ref{sysbouss2}), which depend on variable $v$,
cannot be expressed in terms of $u$ and its derivatives,
since formally $v=D_x^{-1}u_t$.  There is only a finite number of symmetries
of (\ref{sysbouss2}) that do not depend explicitly on the
variable $v$, and only these symmetries do survive the change of variables
to equation (\ref{sysbouss1}). In the extension of the differential
ring by the element $v=D_x^{-1}u_t$ system (\ref{sysbouss1}) has
infinitely many infinitesimal symmetries.

These examples shows that one has to be rather careful with the definition of symmetries and
more rigorous language is indeed required.

\subsection{The ring $\ring$ of differential polynomials.}

In what follows we assume that all functions, such as
$K$ and $G$ in formula (\ref{evsys0}) and (\ref{sys2}),
are differential polynomials of variables $u,v,u_x,v_x,u_{xx}, \ldots$ .
Instead of functions $u,v$ and their partial $x$--derivatives
we introduce an infinite sequence of {\em dynamical variables}
$\{ u_0,v_0,u_1,v_1,u_2,v_2,\cdots \}$ assuming the identification
\begin{equation}\label{dynvars}
 u_0=u,\ v_0=v,\ u_n=\partial^n_x u,\ v_n=\partial^n_x v.
\end{equation}
Often we will omit the zero index and write $u$ and $v$ instead of $u_0$ and $v_0$.

We denote $\ring$ the ring of polynomials over $\bbbc$ of infinite number of dynamical variables.
We assume that $1\not\in \ring$. Elements of the ring $\ring$ are finite sums of monomials with
complex coefficients and therefore each element depends on a finite number of the dynamical variables.
The degree of a monomial is defined as a total power, i.e. the sum of all powers of dynamical
variables that contribute to the monomial. Monomials are eigenfunctions of the following operator
\begin{equation}\label{D0} D_0=X_u+X_v \end{equation}
where
\[ X_u=\sum_{k\ge 0}u_k \frac{\partial}{\partial u_k}\, ,\qquad X_v=\sum_{k\ge 0}v_k \frac{\partial}{\partial v_k}\, .\]
The eigenvalue of $D_0$ is the degree of a monomial. We say that a polynomial is homogeneous
of degree $n$ if every its monomial is of degree $n$. The ring $\ring$ has a natural gradation
\[ \ring = \bigoplus_{n\in \bbbz_+}{\cal R}^n \, ,
\quad {\cal R}^n \cdot {\cal R}^m\subset {\cal R}^{n+m} \, ,
\quad {\cal R}^n=\{ f\in \ring \, |\, D_0 f=nf \}\, .
\]
Elements of ${\cal R}^1$ are linear functions of the dynamical variables, ${\cal R}^2$ quadratic, etc.
It is convenient to define a ``little-oh'' order symbol $o({\cal R}^n)$. We say
that $f=o({\cal R}^n)$ if $f\in \bigoplus_{k>n}{\cal R}^k$, i.e. the degree of every
monomial of $f$ is bigger than $n$.

Let $\mu, \nu$ be two positive integers\footnote{In general,
the weights could be rational numbers including zero, but in this paper we consider
only positive integer weights.}, which we call the weights of $u$ and $v$ respectively
and denote $W(u)=\mu, W(v)=\nu$. We say that a polynomial $f\in\ring$ is a homogeneous
polynomial of weight $\lambda$  (and write $W(f)=\lambda$) if
$ X_{\mu\nu}f=\lambda f$, where
\begin{equation}\label{X}
X_{\mu\nu}=\left(\mu X_u+\nu X_v +X\right)\, , \qquad X=\sum_{k\in
\bbbz_+}k\left( u_k \frac{\partial}{\partial u_k}+v_k
\frac{\partial}{\partial v_k}\right) \, .
\end{equation}
Eigenvalues of the operator $X_{\mu\nu}$ are positive integers and the minimal eigenvalue is equal
to $\mbox{min}(\mu,\nu)$. Monomials are obviously eigenfunctions of $X_{\mu\nu}$.
The {\sl weighted} gradation of $\ring$ is defined as
\[ \ring = \bigoplus_{n\in \bbbz_+}{\cal R}_n \, ,
\quad {\cal R}_n \cdot {\cal R}_m\subset {\cal R}_{n+m} \, ,
\quad {\cal R}_n=\{ f\in \ring \, |\, X_{\mu\nu} f=nf \}\, .
\]

For example if $\mu=1,\nu=2$ then
\[
{\cal R}_1=\bbbc \cdot u_0,\ {\cal R}_2=\mbox{span}(u_0^2, u_1,v_0),\
{\cal R}_3=\mbox{span}(u_0^3,u_0 u_1, u_0 v_0, u_2,v_1),\ldots \, .\]

Operators $D_0$ and $X_{\mu\nu}$ commute and therefore we can define a degree-weighted
gradation of the ring $\ring$
\[ \ring = \bigoplus_{n,p\in \bbbz_+}{\cal R}_p^n \, ,
\quad {\cal R}_p^n \cdot {\cal R}_q^m\subset {\cal R}_{p+q}^{n+m} \, ,
\quad {\cal R}_p^n=\{ f\in \ring \, |\, X_{\mu\nu} f=pf\, ,\, D_{0} f=nf \}\, .
\]

Linear subspaces ${\cal R}^n$ are infinite dimensional, subspaces ${\cal R}_n$ and ${\cal R}_p^n$
are finite dimensional (if we admit zero or negative weights, then the dimension of ${\cal R}_n$
would become infinite).

The ring $\ring$ is a differential ring with derivation
\begin{equation}\label{Dx}
 D_x =\sum _{k\ge 0} \left( u_{k+1} \frac{\partial}{\partial u_{k}}+v_{k+1}\frac{\partial}{\partial v_{k}}\right) \, .
\end{equation}

Derivation $D_x$ represents the $x$--derivative (c.f. (\ref{dynvars})). Since $1\not\in\ring$,
the kernel of the linear map $D_x:\ring\mapsto\mbox{Im}\, D_x\subset\ring$ is empty and
therefore $D_x^{-1}$ is defined uniquely on $\mbox{Im}\, D_x$. It is easy to verify that
\[ [D_0,D_x]=0\, ,\qquad [X_{\mu\nu},D_x]=D_x \]
implying
\[ D_x :{\cal R}_p^n\longmapsto {\cal R}_{p+1}^n\, .\]

With any evolutionary system of equations
\begin{equation}\label{evsys}
u_t =P(u,v,u_x,v_x,\ldots)\, ,\quad v_t =Q(u,v,u_x,v_x,\ldots)\, ,
\end{equation}
where $P,Q$ are differential polynomials, we associate an infinite dimensional dynamical system
\begin{equation}\label{evsysinf}
\frac{d}{dt}(u_k)=D_x^k P(u_0,v_0,u_1,v_1,\ldots),\quad
\frac{d}{dt}(v_k)=D_x^k Q(u_0,v_0,u_1,v_1,\ldots),\qquad  k=0,1,2,\ldots \, .\end{equation}
It is sufficient to define the first two equations ($u=u_0,v=v_0$) of (\ref{evsysinf})
\begin{equation}\label{evsys1}
\frac{d}{dt}u= P,\quad \frac{d}{dt}v=Q\, ,\qquad P,Q\in\ring \, ,
\end{equation}
the rest equations of this infinite system can be obtained by the application of $D_x^k,\ k=1,2,3,\ldots$
assuming that $D_x \frac{d}{dt}=\frac{d}{dt}D_x$.

The system (\ref{evsysinf}) provides a $t$-derivation of the ring $\ring$
\begin{equation}\label{Dtev}
D_t =\sum _{k\ge 0}
\left( D_x^k (P) \frac{\partial}{\partial u_{k}}+D_x^k (Q) \frac{\partial}{\partial v_{k}}\right)\, ,
\end{equation}
The derivations $D_x$ and $D_t$ commute, i.e. $[D_x,D_t]=0$. Derivations of the ring $\ring$ that
commute with $D_x$ we call {\sl evolutionary derivations}. There is one to one correspondence between
evolutionary systems of PDEs and evolutionary derivations. We say that evolutionary derivation $D_t$ is generated by
$(P,Q)$, therefore it is convenient to adopt notation $D_{\bf a}$ where ${\bf a}=(P,Q)^{\mbox{tr}}$
denotes a vector column  of the right hand side of (\ref{evsys1}).

The evolutionary system (\ref{evsys}) and corresponding derivation $D_t$ are called homogeneous of
weight $s$ and write $W(D_t)=s$ if $P,Q$ are homogeneous elements of $\ring$ and $W(P)-\mu=W(Q)-\nu=s$.
If $W(D_t)=s$ then
\[ D_t :{\cal R}_n \longmapsto {\cal R}_{n+s}\, .\]
Moreover, if $P\in {\cal R}_{s+\mu}^k,Q\in {\cal R}_{s+\nu}^k$ then
$D_t:{\cal R}_n^m\longmapsto {\cal R}_{n+s}^{m+k-1}$.

For example $D_0$, cf. (\ref{D0}) and $D_x$, cf. (\ref{Dx}) are a homogeneous derivations of $\ring$
generated by $(u_0,v_0)^{\mbox{tr}}$ and $(u_1,v_1)^{\mbox{tr}}$ respectively with $W(D_0)=0, W(D_x)=1$,
while $X$ (\ref{X}) is a non-evolutionary derivation of $\ring$.

It follows from (\ref{Dtev}) that for any element $g\in\ring$ we have
\[ D_t g=\sum _{k\ge 0}
\left( D_x^k (P) \frac{\partial g}{\partial u_{k}}+D_x^k (Q)
\frac{\partial g}{\partial v_{k}}\right)= g_{*u}(P)+g_{*v}(Q) \]
where
\begin{equation}\label{g*uv}
g_{*u}=\sum_{k\ge 0}\frac{\partial g}{\partial u_k}D_x^k\, ,\qquad
g_{*v}=\sum_{k\ge 0}\frac{\partial g}{\partial v_k}D_x^k
\end{equation}
are two linear differential operators. The sums in (\ref{g*uv}) are finite, since $g$ depends on a
finite number of dynamical variables. Thus, with any $g\in\ring$ we associate two linear operators, a vector row
\begin{equation}\label{fder} g\to g_*=(g_{*u}\, ,\ g_{*v})\ , \end{equation}
which is called the Fr\'echet derivative of $g$.

Let us consider a two dimensional liner space ${\cal
K}=\ring$ over $\bbbc$  of vector columns whose entries
are elements of $\ring$. Every element ${\bf
a}=(P,Q)^{\mbox{tr}}\in\cK$ generates an evolutionary derivation
$D_{\bf a}$  ($D_t$ cf. formula (\ref{Dtev})) of the ring and vice
versa, with any evolutionary derivation we associate a generator,
i.e. an element of $\cK$. Derivations of $\ring$ form a Lie algebra
over $\bbbc$ with the Lie product of two derivations defined as a
commutator of the derivations. The evolutionary derivations form a
subalgebra of the Lie algebra of all derivations. Indeed, it is easy
to verify that if $D_{\bf a}$ and $D_{\bf b}$ are two evolutionary
derivations generated by ${\bf a}$ and ${\bf b}$ respectively, then
the commutator $D_{\bf a}\circ D_{\bf b}-D_{\bf b}\circ D_{\bf a}$
is an evolutionary derivation $D_{[{\bf a},{\bf b}]}$ generated by
the Lie bracket $[{\bf a},{\bf b}]\in\cK$ of elements ${\bf a}$ and
${\bf b}$ where
\begin{equation}\label{liebrac}
[{\bf a},{\bf b}]={\bf b}_*({\bf a})- {\bf a}_*({\bf b}),
\end{equation}
or in components, assuming ${\bf a}=
\left(\begin{array}{c}P\\Q\end{array}\right)\, ,\ {\bf b}=\left(\begin{array}{c}S\\T\end{array}\right)$
we have
\begin{equation}\label{lieD}
 [{\bf a},{\bf b}]=\left(\begin{array}{c}
S_{*u}(P)+S_{*v}(Q)-P_{*u}(S)-P_{*v}(T)\\
T_{*u}(P)+T_{*v}(Q)-Q_{*u}(S)-Q_{*v}(T)
\end{array}\right)\, .
\end{equation}
In particular $[D_{\bf a},D_{\bf b}]=0\, \Longleftrightarrow \, [{\bf a},{\bf b}]=0$.
We shall assume that the linear space $\cK$ is equipped with the Lie bracket (\ref{liebrac})
and therefore it is a Lie algebra.

The components of the Fr\'echet derivative are examples of
differential operators with coefficients from  the ring $\ring$
extended by the base field $\bbbc$:
\[ \ring_\bbbc=\bbbc\oplus\ring.\]
We define formal series of the form
\begin{equation}\label{fser}
A=\sum_{k\ge 0} a_{N-k} D_x^{N-k}\, ,\quad a_m\in \ring_\bbbc\, ,\quad m\in \bbbz \, .
\end{equation}
The order of the formal series (\ref{fser}) is $N$ (we assume that the leading coefficient $a_N\neq 0$).
The positive part of $A$, denoted by  $A_+$, which is a finite sum of the form $A_+=\sum_{k=0}^N a_{N-k} D_x^{N-k}$
is a differential operator with obvious action on $\ring$, while a formal series is not an operator,
since no action on the ring $\ring$ is defined. The sum of formal series is defined obviously,
a multiplication (composition) is defined by
\begin{equation}\label{fscomp}
 a_n D_x^n \circ b_m D_x^m=\sum _{k\ge 0}\left(\begin{array}{c}n\\k\end{array}\right)
a_n D_x^k (b_m)D_x^{m+n-k} \, .
\end{equation}
For positive $n$ the sum (\ref{fscomp}) is finite since the binomial coefficients
\[  \left(\begin{array}{c}n\\k\end{array}\right)=
\frac{n(n-1)(n-2)\cdots (n-k+1)}{k!} \]
vanish for $k>n$, for negative $n$ the composition is well defined in the sense of formal series.
Formal series form an associative ring and we denote it by
 $\ring_\bbbc [D_x]$.

Suppose $A$ is a formal series (differential operator) of the form (\ref{fser}),
then the conjugated formal series (differential operator) is defined as
\begin{equation}\label{conjA}
 A^\dagger=\sum_{k\ge 0} (-1)^{N-k} D_x^{N-k}\circ a_{N-k} \, .
\end{equation}

We also consider formal series and differential operators with
matrix coefficients, assuming that the entries of the matrix
belong to $\ring_\bbbc$ (the Fr\'echet derivative
(\ref{fder}) is an example of the operator with vector
coefficients). In this case  the coefficients of the conjugated
series (\ref{conjA}) have to be transposed.

For any $g\in\ring$ we define a vector column of variational derivatives $\delta g$
\begin{equation}\label{varder} \delta g=\left(\begin{array}{c}\delta_u g\\
\delta_v g\end{array}\right)=g_*^\dagger(1)=\left(\begin{array}{c} g_{*\, u}^\dagger (1)\\
g_{*\, v}^\dagger (1)\end{array}\right)
\end{equation}
There is a useful statement \cite{gmsh}, that $\delta g=0$ if and only if $g\in\mbox{Im}\, D_x$,
i.e. there exists $h\in\ring$ such that $g=D_x (h)$. Thus, the variational derivative is well
defined on the factor space $\ring/\mbox{Im}\, D_x$, indeed two elements $a,b\in\ring$ have
the same variational derivative iff their difference is a total derivative, that is, $a-b\in \mbox{Im}\, D_x$.

\subsection{Symmetries.}

There are many equivalent definitions of infinitesimal symmetries of
PDEs (based on infinitesimal transformations of solutions, commuting
vector fields, etc., see for example \cite{mss,ss}). In this
paper we adopt a definition, which is quite convenient for practical
computations as well as for theoretical considerations.

Suppose we have an evolutionary system (\ref{evsys1}) and the column vector of
the right-hand side is ${\bf a}=(P,Q)^{\mbox{tr}}$. We say that ${\bf b}\in\cK$ is
a symmetry (a generator of an infinitesimal symmetry) if $[{\bf a},{\bf b}]=0$.
The evolutionary system corresponding to ${\bf b}$ is compatible with (\ref{evsys1})
and often it is called symmetry as well. Due to the Jacobi identity a commutator of
two symmetries is again a symmetry, therefore symmetries form a subalgebra of the Lie algebra $\cK$.

Let us consider equation (\ref{eqnon}) which is represented by evolutionary system (\ref{evsys0}).
In this case ${\bf a}=(v,K)^{\mbox{tr}}$. Assuming
 ${\bf b}=(S,F)^{\mbox{tr}}$, and evaluating $[{\bf a},{\bf b}]=0$ using (\ref{lieD}), we receive
\[
S_{*\, u}(v_0)+S_{*\, v}(K)=F\, ,\quad
F_{*\, u}(v_0)+F_{*\, v}(K)=K_{*\, u}(S)+K_{*\, v}(F)\, ,
\]
which can be rewritten as
\[ D_{t}(S)=F\, ,\quad D_t (F)=D_{\bf b}(K)\, , \]
where
\begin{equation}\label{Dt0} D_t =\sum _{k\ge 0} \left( v_{k} \frac{\partial}{\partial u_{k}}+
D_x^k (K) \frac{\partial}{\partial v_{k}}\right)\, ,\quad
D_{\bf b} =\sum _{k\ge 0} \left( D_x^k(S) \frac{\partial}{\partial u_{k}}+
D_x^k D_t(S) \frac{\partial}{\partial v_{k}}\right)
\end{equation}
A symmetry generator ${\bf b}$  is completely determined by its first component $S$, which satisfies equation
\begin{equation}\label{symeq0}
D_t^2 (S)=D_{\bf b}(K)\, .
\end{equation}
Equation (\ref{symeq0}) together with (\ref{Dt0}) can be taken as a definition of symmetry
for evolutionary system (\ref{evsys0}).

Evolutionary system (\ref{sys2}) represents equation (\ref{eqnon})
in the standard set of dynamical variables (\ref{dynvars}) if
function $K=D_x(G)$ does not depend on the variable $v$ (it may of
course depend on $v_x=u_t, v_{xx}=u_{tx},\ldots $ ). Indeed, the
variable $v$ cannot be represented in terms of $u$ and its
derivatives. Its symmetries may explicitly depend on $v$ and
therefore the corresponding infinitesimal transformations are
non-local, i.e. cannot be expressed in terms of $u$ and its
derivatives. System (\ref{sysbouss2}) has infinitely many
symmetries, but only a finite number of them does not depend on $v$
explicitly. Thus, only a finite number of symmetries of the
Boussinesq equation are local. The meaning of the notion "locality"
depends on the choice of the set of dynamical variables.
Representing the Boussinesq in the form (\ref{sys2}) we introduce a
variable $v$ and extend the standard set. In this extended set of
dynamical variables the Boussinesq equation (\ref{bouss1}) has
infinitely many local infinitesimal symmetries. In the case of the
system (\ref{sys2}), there is another natural extension of the
standard set of dynamical variables (\ref{dynvars}). Due to the
structure of the first equation $u_t=v_1$ we can introduce a
``potential'' $w$ such that $u=w_1,u_1=w_2,\ldots u_k=w_{k+1}$. In
this new extended set symmetries that depend explicitly on the
variable $v$ become local ($v=w_t$). Thus, the potential version of
the Boussinesq equation
\begin{equation}\label{potbouss}
w_{tt}=w_4+2 w_1 w_2
\end{equation}
has infinitely many local symmetries. In the case of the Boussinesq equation one can extend
the standard set one step further, introducing the next potential $z$ such that
$z_1=w,z_2=w_1=u,\ldots z_{n+2}=w_{n+1}=u_n$ and therefore $z_{tt}=z_4+(z_2)^2$.
In the case of homogeneous equations introduction of a potential reduces the weight of
the variable by one (in the case of the Boussinesq equation (\ref{bouss1}) we have $W(u)=2$,
for the potential Boussinesq equation (\ref{potbouss})  $W(w)=1$, and $W(z)=0$).

For system (\ref{sys2}) we have ${\bf a}=(v_1,G)^{\mbox{tr}}$.
Denoting ${\bf b}=(S,F)^{\mbox{tr}}$ it is easy to show that
condition $[{\bf a},{\bf b}]=0$ can be written as
\begin{equation}\label{symsys1}
 D_{t}(S)=D_x(F)\, ,\quad D_t (F)=D_{\bf b}(G)\, ,
\end{equation}
where
\begin{equation}\label{Dt1} D_t =\sum _{k\ge 0} \left( v_{k+1}
\frac{\partial}{\partial u_{k}}+D_x^k (G) \frac{\partial}{\partial v_{k}}\right)\, ,\quad
D_{\bf b} =\sum _{k\ge 0} \left( D_x^k(S) \frac{\partial}{\partial u_{k}}+
D_x^{k-1}D_t(S) \frac{\partial}{\partial v_{k}}\right)\, .
\end{equation}
It follows from the first equation of the system (\ref{symsys1}) that $S$ is either a
density of a local conservation law of the system (\ref{sys2}) or a total derivative ($S\in\mbox{Im}\, D_x$).
In both cases $F=D_x^{-1}D_t(S)\in \ring$ and $F$ can be found uniquely if $S$ is known.
Eliminating $F$ from (\ref{symsys1}) we arrive to equation $D_t^2 (S)=D_{\bf b}(K),\, K=D_x (G)$,
which looks exactly as (\ref{symeq0}), but the difference is in the definition of derivations (\ref{Dt1}).

\begin{Def} We say that differential polynomial $S\in \ring$ is a generator of {\em approximate}
symmetry of {\em degree} $n$ of system (\ref{sys2}) if
\[ D_t^2 S-D_{\bf b} K=o({\cal R}^n) \]
($D_t$ and $D_{\bf b}$ are given by (\ref{Dt1})). We say that system (\ref{sys2})
has an infinite hierarchy of approximate symmetries of degree $n$ if every member
of the hierarchy is an approximate symmetry of degree $n$.
\end{Def}

Approximate symmetries of degree $1$ are simply the symmetries of the linear part of the system.
Any system (\ref{sys2}) has infinitely many approximate symmetries of degree $1$.
The requirement of the existence of approximate symmetries of degree $2$ is very restrictive
and highly non-trivial. An equation may have infinitely many approximate symmetries of degree $2$,
but fail to have approximate symmetries of degree $3$. Integrable equations have infinite
hierarchies of approximate symmetries of any degree. The degree of the hierarchy of approximate
symmetries is, in a certain sense, a measure of integrability.

A homogeneous equation has homogeneous symmetries. Indeed, suppose
$S$ generates a symmetry and is a sum of components
$S=S_1+\cdots +S_k$ of different weights $W(S_p)=n_p$, then it
is easy to check that each homogeneous component $S_p$ generates a
symmetry. We define the order of a homogeneous symmetry $S$ as the
weight of the corresponding evolutionary derivation $\ord (S)=
W(D_{\bf b})=W(S)-W(u)$, which is obviously homogeneous. For example
$\ord (u_1)=W(D_x)=1$, for systems of equations  (\ref{sysbouss1}),
(\ref{sysbouss2}) and (\ref{sysbouss3}) we have $\ord
(u_t)=W(D_t)=2$. Therefore the order of a homogeneous
non-evolutionary equation of the form (\ref{eqnon}) is natural to
define as $2W(D_t)$. Thus the order of the Boussinesq equation
(\ref{bouss1}) is 4, the order of equation (\ref{shabclass}) is 3.

\subsection{Recursion, Hamiltonian and symplectic operators}\label{RHS}

Integrability of an evolutionary equation we associate with the existence of an
infinite hierarchy of its symmetries. If a recursion operator is known, then symmetries
can be found recursively and therefore the existence of a recursion operator is a
sufficient condition for integrability.

For evolutionary system (\ref{evsys}) a recursion operator can be defined as a
"pseudo-differential operator" $R$ with coefficients in $\ring$ which satisfies
the following operator equation
\begin{equation}\label{recop}
D_{\bf a} (R)={\bf a}_* \circ R-R\circ {\bf a}_* \, ,
\end{equation}
where ${\bf a}_* $ is a Fr\'echet derivative of vector ${\bf
a}=(P,Q)^{\tr}$. If action of $R$ is well defined on a symmetry ${\bf
b}_1$, i.e. ${\bf b_2}=R({\bf b}_1)\in \cK$, then
${\bf b}_2$ is a new symmetry of  the evolutionary system
(\ref{evsys}). Starting from a ``seed'' symmetry ${\bf b}_1$, one
can build up an infinite hierarchy of symmetries ${\bf
b}_{n+1}=R^n({\bf b}_1)$, provided that each action of $R$ produces
an element of the ring $\ring$, i.e. $R^n ({\bf b}_1)\in \cK$.

For example, if we represent the Boussinesq equation (\ref{bouss1})
in the form of evolutionary system (\ref{sysbouss2}), then
\[ {\bf a}_*=  \left(\begin{array}{cc}
0&D_x\\ D_x^3+2uD_x+2u_1&0
                     \end{array}\right) \]
and it is easy to verify that a pseudo-differential operator
\begin{equation}\label{recboussR}
R=\left(\begin{array}{cc}
3v+2v_1 D_x^{-1}& 4D_x^{2}+2u+u_1D_x^{-1}\\
4D_x^4+10uD_x^2+15u_1D_x+9u_2+4u^2+(2u_3+4uu_1)D_x^{-1} & 3v+v_1 D_x^{-1}
\end{array}
\right)
\end{equation}
satisfies equation (\ref{recop}) and therefore is a recursion operator.

In the previous section we have shown that the second component of
a symmetry is completely determined by its first component. We can restrict the action
of the recursion operator on the first component. If $S$ is the first component of a symmetry
of the Boussinesq equation (\ref{sysbouss2}), then
\[ \hat{S}=\Re(S)=\left( 3v+2v_1 D_x^{-1}+(4D_x^{2}+2u+u_1D_x^{-1})D_t D_x^{-1}\right) (S) \]
is the first component of the next symmetry in the hierarchy. We call $\Re$ a restricted recursion operator.
The restricted recursion operator $\Re$ of the Boussinesq equation is a
homogeneous pseudo-differential operator of weight $W(\Re)=3$ (the weight $W(D_x^{-1})=-1$),
therefore $\ord (\hat{S})=\ord (S)+3$.

A space-shift, generated by ${\bf b}_1=(u_1,v_1)^{\tr}$, is a symmetry of the
Boussinesq equation (\ref{sysbouss2}). Taking $u_1$ as a seed, we can construct
an infinite hierarchy $S_{3k+1}=\Re^k(S_1)$ of
symmetries of orders $3k+1, k=0,1,2,\ldots$. For example
\[ S_4=4v_3+4v_1 u+4 vu_1=4D_x (v_2+vu)\, . \]
We see that $S_4$ is a total derivative and therefore
$S_7=\Re(S_4)\in \ring$ is the next symmetry in the hierarchy, etc.
The Boussinesq equation itself is not a member of this hierarchy. If we
take a seed symmetry, corresponding to the time-shift ${\bf
c}_1=(v_1,u_3+2uu_1)^{\tr}$ we receive another infinite hierarchy of
symmetries $S_{3k+2}=\Re^k(v_1), k=0,1,2,\ldots$. The Boussinesq
equation does not have symmetries of order $3k, k\in\mathbb N$. One
can show that $S_{3k+1}$ and $S_{3k+2}$ are elements of the ring
$\ring$ for any $k\in \mathbb N$ and therefore $\Re$ generates two
infinite hierarchies of symmetries of the Boussinesq equation.
Moreover, all symmetries from the both hierarchies commute with each
other.

For system (\ref{sys2}) a recursion operator is completely determined by its two
entries $R_{11}$ and $R_{12}$. Indeed, it follows from (\ref{recop}) that
\begin{equation}\label{R21R22}
R_{21}=D_x^{-1}\circ (D_t (R_{11})+R_{12}\circ G_{*u})\, ,\quad
R_{22}=D_x^{-1}\circ (D_t (R_{12})+R_{11}\circ D_x+R_{12}\circ G_{*v})\, .
\end{equation}
The restricted recursion operator $\Re$ for system (\ref{sys2}) can be represented as
\begin{equation}\label{restrR}
\Re=R_{11}+R_{12}D_t D_x^{-1}\, .
\end{equation}

If the system of equations is bi-Hamiltonian, then there is a useful
splitting of the recursion operator in a composition $R=\cH \circ \cJ$
of a Hamiltonian $\cH$ and symplectic $\cJ$ operators. A
comprehensive definition and theory of these operators one can find
in the fundamental monograph of Irene Dorfman \cite{dorfman}. Here
we remind basic facts and introduce notations.

A "pseudo-differential operator" $\cH$ is called a Hamiltonian
operator if it is skew-symmetric $\cH^\dagger=-\cH$ and defines a
Poisson bracket for any two elements $f,g\in \ring$:
\[ \{f,g\}=\left(\delta f^{\tr}\cdot\cH(\delta g)\right) , \]
which satisfy the Jacobi identity. A Hamiltonian operator $\cH$ of an evolutionary system
(\ref{evsys}) should also be invariant (its Lie derivative should vanish):
\begin{equation}\label{eqcH}
D_{\bf a}\cH={\bf a}_*\circ \cH +\cH\circ {\bf a}_*^\dagger\, .
\end{equation}
An evolutionary system (\ref{evsys}) is called Hamiltonian, if  ${\bf a}=\cH (\delta H)$,
where $H$ is a density of a conservation law. The density $H$ is called the Hamiltonian of the system.
If $\rho$ is a density of a conservation law of a Hamiltonian system (\ref{evsys}),
then ${\bf b}=\cH(\delta \rho)$ is a generator of a symmetry of the system.

A symplectic operator $\cJ$ is, in a sense,  inverse to a Hamiltonian operator.
It is also skew-symmetric $\cJ^\dagger=-\cJ $ and defines a closed 2-form.
It should also  be invariant (its Lie derivative should vanish):
\begin{equation}\label{eqcJ}
D_{\bf a}\cJ=-{\bf a}_*^\dagger \circ \cJ -\cJ\circ {\bf a}_*\, .
\end{equation}

It follows from (\ref{eqcH}), (\ref{eqcJ}) that the composition $\cH\circ \cJ$ satisfy equation
(\ref{recop}). Indeed,
\[ D_{\bf a}(\cH\circ \cJ)=D_{\bf a}(\cH)\circ  \cJ+\cH\circ D_{\bf a}(\cJ)=
{\bf a}_*\circ \cH\circ\cJ +\cH\circ {\bf a}_*^\dagger\circ\cJ -
\cH\circ{\bf a}_*^\dagger \circ \cJ -\cH\circ\cJ\circ {\bf a}_*\, .
\]

In the case of evolutionary system (\ref{sys2}) a Hamiltonian
operator $\cH$ is $2\times 2$ matrix whose entries $\cH_{ij}$ are
differential (or pseudo-differential) operators. It follows from
(\ref{eqcH}) that
\begin{equation}\label{cH22}
\cH_{22}=G_{*u}\circ \cH_{11}\circ D_x^{-1}+G_{*v}\circ \cH_{21}\circ D_x^{-1}-D_{\bf a}(\cH_{21})\circ D_x^{-1}
\end{equation}
and $\cH_{12}=-\cH_{21}^\dagger$  due to the skew-symmetry.
Thus two entries $\cH_{11}$ and $\cH_{21}$ completely determine the Hamiltonian operator.

Similarly for a symplectic operator we find from (\ref{eqcJ}) that
 \begin{equation}\label{cJ11}
\cJ_{11}=-G_{*u}^\dagger \circ \cJ_{22}\circ D_x^{-1}- \cJ_{12}\circ G_{*v}\circ D_x^{-1}-D_{\bf a}(\cJ_{12})\circ D_x^{-1}
\end{equation}
and  $\cJ_{21}=-\cJ_{12}^\dagger$. Two components $\cJ_{22}$ and
$\cJ_{12}$ completely determine the symplectic operator  of system
(\ref{sys2}).

A restricted recursion operator for equation (\ref{sys2}) can be represented in the form
\begin{equation}\label{recopHJ}
\Re={\cal
H}_{11}\circ  \cJ_{11} +{\cal H}_{12}\circ  \cJ_{21} +({\cal H}_{11}\circ
\cJ_{12} +{\cal H}_{12}\circ  \cJ_{22})\circ D_x^{-1} D_t \, .
\end{equation}

In the case of the Boussinesq equation (\ref{sysbouss2}) there are
Hamiltonian and symplectic operators of the form
\[ \cH=\left(\begin{array}{cc}
3D_x^3+uD_x+D_x u&3vD_x+2v_1\\
3vD_x+v_1& 4D_x^5+5(uD_x^3+D_x^3 u)-3 (u_2D_x+D_x u_2)+4uD_x u
\end{array}\right)\, ,\quad
\cJ=\left(\begin{array}{cc}
0&D_x^{-1}\\
D_x^{-1}&0
\end{array}\right)\, .\]
It is easy to check that  $\cJ^{-1}$ is also a Hamiltonian operator for the Boussinesq equation and compatible with $\cH$.

\subsection{Symbolic Representation}

In this section we remind basic definitions and notations of perturbative symmetry
approach in symbolic representation (for more details see e.g. \cite{mn, jp}).

We start with the symbolic representation $\hat{\cal R}$ of the ring
$\ring$. Linear monomials $u_n, v_m$  are represented by
\begin{equation}\label{unvm}
u_n\to u\xi_1^n,\quad v_m\to v\zeta_1^m
\end{equation}
Quadratic monomials $u_nu_m$, $u_nv_m$, $v_nv_m$  have the following symbols
\begin{equation}\label{monom}
u_nu_m\to\frac{u^2}{2}(\xi_1^n\xi_2^m+\xi_1^m\xi_2^n),\quad u_nv_m\to uv(\xi_1^n\zeta_1^m),\quad v_nv_m\to \frac{v^2}{2}
(\zeta_1^n\zeta_2^m+\zeta_1^m\zeta_2^n)
\end{equation}
A symbolic representation of a monomial $u_0^{n_0}u_1^{n_1}\cdots
u_p^{n_p}v_0^{m_0}v_1^{m_1}\cdots v_q^{m_q},\, n_0+n_1+\cdots
+n_p=n,\, m_0+m_1+\cdots+m_q=m$ is defined as:
\begin{eqnarray}\nonumber
 &&u_0^{n_0}u_1^{n_1}\cdots
u_p^{n_p}v_0^{m_0}v_1^{m_1}\cdots v_q^{m_q}\to\\ \label{mon} &&\to
u^nv^m
\langle\xi_1^0\xi_2^0\cdots\xi_{n_0}^0\xi_{n_0+1}^1\cdots
\xi_{n_0+n_1}^1\cdots\xi_n^p\rangle_{\xi}\langle\zeta_1^0\zeta_2^0\cdots\zeta_{m_0}^0\zeta_{m_0+1}^1
\cdots\zeta_{m_0+m_1}^1\cdots\zeta_m^q\rangle_{\zeta}\ ,
\end{eqnarray}
where triangular brackets $\langle\rangle_{\xi}$ and
$\langle\rangle_{\zeta}$ denote the averaging over the group
of permutations of $n$ elements $\xi_1,\ldots,\xi_n$, denoted
by $\Sigma_n$, i.e.
\[
\langle c(\xi_1,\ldots,\xi_n,\zeta_1,\ldots
,\zeta_m)\rangle_{\xi}=\frac{1}{n!}\sum_{\sigma\in\Sigma_n}c(\sigma(\xi_1),\ldots,\sigma(\xi_n),\zeta_1,\ldots
,\zeta_m)
\]
and $m$ elements $\zeta_1,\ldots,\zeta_m$ respectively.

To define the symbolic representation $\hat{\ring}$ of the ring
$\ring$ we need to define summation, multiplication and derivation
in $\hat{\ring}$. To the sum of two elements of the ring corresponds
the sum of their symbols. To the product of two elements
$f,g\in\ring$ with symbols $f\to
u^nv^ma(\xi_1,\ldots,\xi_n,\zeta_1,\ldots,\zeta_m)$ and $g\to
u^pv^qb(\xi_1,\ldots,\xi_p,\zeta_1,\ldots,\zeta_q)$ corresponds:

\begin{equation}\label{multmonoms}
fg\to u^{n+p}v^{m+q}\langle\langle
a(\xi_1,\ldots,\xi_n,\zeta_1,\ldots,\zeta_m)b(\xi_{n+1},\ldots,\xi_{n+p},
\zeta_{m+1},\ldots,\zeta_{m+q})\rangle_{\xi}\rangle_{\zeta},
\end{equation} where the symmetrisation operations are taken with respect to
permutations of all arguments $\xi$ and $\zeta$. It is easy to see
that representations of quadratic (\ref{monom}) and general
(\ref{mon}) monomials immediately follows from (\ref{unvm}) and
(\ref{multmonoms}).

If $f\in\ring$ has a symbol $f\to
u^nv^ma(\xi_1,\ldots,\xi_n,\zeta_1,\ldots,\zeta_m)$, then the
symbolic representation for its $N$-th derivative $D_x^N(f)$ is:
\[
D_x^N(f)\to
u^nv^m(\xi_1+\xi_2+\cdots+\xi_n+\zeta_1+\zeta_2+\cdots\zeta_m)^N
a(\xi_1,\ldots,\xi_n,\zeta_1,\ldots,\zeta_m)
\]

Thus we obtained the symbolic representation $\hat{\cal R}$ of the
differential ring $\ring$.

To construct the symbolic representation $\hat{\ring} [\eta]$ of the
associative ring of formal series $\ring_\bbbc [D_x]$ we define a
symbol $\eta$ which corresponds to the derivation operator $D_x$ and
satisfies the following rules of action and composition:
\begin{eqnarray}
\nonumber
&&\eta(u^nv^ma(\xi_1,\ldots,\xi_n,\zeta_1,\ldots,\zeta_m))=\\ \nonumber
&&=u^nv^m(\xi_1+\xi_2+\cdots+\xi_n+\zeta_1+\zeta_2+\cdots\zeta_m)a(\xi_1,\ldots,\xi_n,\zeta_1,\ldots,\zeta_m),
\\ \nonumber
&&\eta\circ u^nv^ma(\xi_1,\ldots,\xi_n,\zeta_1,\ldots,\zeta_m)=\\ \nonumber
&&=u^nv^m(\xi_1+\xi_2+\cdots+\xi_n+\zeta_1+\zeta_2+\cdots\zeta_m+\eta)a(\xi_1,\ldots,\xi_n,\zeta_1,\ldots,\zeta_m)
\end{eqnarray}
The last expression represents the Leibnitz rule
 $D_x\circ f=D_x(f)+f\circ D_x$ in the symbolic representation. Moreover, for any $N\in\bbbz$
\[ D_x^N\circ f\to \eta^p\circ u^nv^ma(\xi_1,\ldots,\xi_n,\zeta_1,\ldots,\zeta_m)=
 u^nv^m(\xi_1+\xi_2+\cdots+\xi_n+\zeta_1+\zeta_2+\cdots\zeta_m+\eta)^N a(\xi_1,\ldots,\xi_n,\zeta_1,\ldots,\zeta_m).\]

For any two terms $fD_x^p,\,gD_x^q$ of formal series ($p,q\in{\bbbz}$) with symbols  $f\to
u^nv^ma(\xi_1,\ldots,\xi_n,\zeta_1,\ldots,\zeta_m),\ g\to
u^sv^rb(\xi_1,\ldots,\xi_s,\zeta_1,\ldots,\zeta_r)$ the composition rule in the symbolic representation reads
\begin{eqnarray}
\nonumber &&fD_x^p\circ gD_x^q\to\\ \nonumber
&&u^{n+s}v^{m+r}\langle\langle
a(\xi_1,\ldots,\xi_n,\zeta_1,\ldots,\zeta_m)(\xi_{n+1}+\xi_{n+2}+\cdots+\xi_{n+s}+\zeta_{m+1}+\zeta_{m+2}+
\cdots+\zeta_{m+r}+\eta)^p\times\\ \label{comprulepq} &&\times
b(\xi_{n+1},\xi_{n+2},\ldots,\xi_{n+s},\zeta_{m+1},\zeta_{m+2},\ldots,\zeta_{m+r})\eta^q\rangle_{\xi}\rangle_{\zeta},
\end{eqnarray}
where the symmetrisation is taken with respect to permutations of arguments $\xi$ and arguments $\zeta$,
 but not the argument $\eta$.

Elements of $\hat{\ring} [\eta]$ are formal series
\begin{equation}
\label{A}
A=a_{00}(\eta)+ua_{10}(\xi_1,\eta)+va_{01}(\zeta_1,\eta)+u^2a_{20}(\xi_1,\xi_2,\eta)+
uva_{11}(\xi_1,\zeta_1,\eta)+v^2a_{02}(\zeta_1,\zeta_2,\eta)+u^3a_{30}(\xi_1,\xi_2,\xi_3,\eta)+\cdots,
\end{equation}
where functions
$a_{nm}(\xi_1,\ldots,\xi_n,\zeta_1,\ldots,\zeta_m,\eta)$ are symmetric
functions with respect to permutations of arguments $\xi_i$ and
arguments  $\zeta _i$. Functions
$a_{nm}(\xi_1,\ldots,\xi_n,\zeta_1,\ldots,\zeta_m,\eta)$ are not
necessarily polynomials, but they possess an important property
which we call locality:

\begin{Def}\label{localf}
We shall call a function
$a_{nm}(\xi_1,\ldots,\xi_n,\zeta_1,\ldots,\zeta_m,\eta)$ local if all
the coefficients
of its expansion
\[
a_{nm}(\xi_1,\ldots,\xi_n,\zeta_1,\ldots,\zeta_m,\eta)=\sum_{k\le s}
a_{nm}^k(\xi_1,\ldots,\xi_n,\zeta_1,\ldots,\zeta_m)\eta^{k},\quad
\eta\to\infty
\]
are polynomials in variables
$\xi_1,\ldots,\xi_n,\zeta_1,\ldots,\zeta_m$.
\end{Def}

The property of locality reflects the fact that coefficients of a
formal series $A\in\ring[D_x]$ are differential polynomials, i.e.
elements of the ring $\ring_\bbbc$.

Formal series (\ref{A}) form a ring: the summation is evident,
while the multiplication (composition) rule is given by (compare
with (\ref{comprulepq})):
\begin{eqnarray}
\nonumber
&&u^nv^ma_{nm}(\xi_1,\ldots,\xi_n,\zeta_1,\ldots,\zeta_m,\eta)\circ
u^pv^rb_{pr}(\xi_1,\ldots,\xi_p,\zeta_1,\ldots,\zeta_r,\eta)=\\ \nonumber
&&=u^{n+p}v^{m+r}\langle\langle a_{nm}(\xi_1,\ldots,\xi_n,\zeta_1,\ldots,
\zeta_m,\eta+\xi_{n+1}+\cdots+\xi_{n+p}+\zeta_{m+1}+\cdots+\zeta_{m+r})\times \\
\nonumber
&&\times b_{pr}(\xi_{n+1},\ldots,\xi_{n+p},\zeta_{m+1},\ldots,\zeta_{m+r},\eta)\rangle_{\xi}\rangle_{\zeta} .
\end{eqnarray}

The ring $\hat{\ring}[\eta]$ is graded
\[ \hat{\ring}[\eta]=\bigoplus_{n=0}\hat{\ring}^n[\eta], \]
where $\hat{\ring}^n[\eta]$ with $n=0,1,2,$ etc. are constant (i.e.
$u,v$ independent), linear in $u$ or $v$, quadratic, cubic,  etc.
In other words, $\hat{\ring}^n[\eta]$ is eigenspace of
operator $D_0$, cf. (\ref{D0}) corresponding to eigenvalue $n$. We say
that a formal series $A=o(\hat{\ring}^n[\eta])$ if $A\in
\bigoplus_{k>n}\hat{\ring}^k[\eta]$.

\subsection{Formal Recursion Operator}

Choosing symbolic representation we make calculus of derivations
much simpler paying the price of  a complicated multiplication rule
for symbols. For example, the symbolic representation of the
Fr\'echet derivative $f_*=(f_{*,u},f_{*,v})$ of an element $f\in
\ring$ with a symbol
$u^nv^ma(\xi_1,\ldots,\xi_n,\zeta_1,\ldots,\zeta_m)$
 is:
\begin{equation}\label{frechet}
(f_{*,u},f_{*,v})\to (
nu^{n-1}v^ma(\xi_1,\ldots,\xi_{n-1},\eta,\zeta_1,\ldots,\zeta_m),\,
mu^nv^{m-1}a(\xi_1,\ldots,\xi_{n},\zeta_1,\ldots,\zeta_{m-1},\eta))\,
.
\end{equation}
The variational derivative (\ref{varder}) of $f$ in symbolic
representation takes form
\[
\delta f\to\left(
nu^{n-1}v^ma(\xi_1,\ldots,\xi_{n-1},-\sum_{k=1}^{n-1}\xi_k,\zeta_1,\ldots,\zeta_m),
mu^nv^{m-1}a(\xi_1,\ldots,\xi_{n},\zeta_1,\ldots,\zeta_{m-1},-\sum_{k=1}^{m-1}\zeta_k)\right)\,
.
\]
In symbolic representation the problem of computation of symmetries
and conservation laws can be reduced to linear algebra \cite{jp,mn}.

We start with  system (\ref{sys2}) where $G\in\ring$ and it can be represented as
\[
G=G_1 +G_2 +G_3 +\cdots,\quad G_n \in\ring^n\, ,
\]
and the corresponding evolutionary derivation is generated by the
vector ${\bf a} = (v_1, G) ^ {\mbox{tr}}$.

Its symbolic representation is of the form
\begin{eqnarray}
\left\{\begin{array}{l}
\label{syssym} u_t=v\zeta_1\\
v_t=u\omega_1(\xi_1)+v\omega_2(\zeta_1)+u^2a_{20}(\xi_1,\xi_2)+u v
a_{11}(\xi_1,\zeta_1)+v^2a_{02}(\zeta_1,\zeta_2)+u^3a_{30}(\xi_1,\xi_2,\xi_3)+\\
\quad +u^2va_{21}(\xi_1,\xi_2,\zeta_1)+
uv^2a_{12}(\xi_1,\zeta_1,\zeta_2)+v^3a_{03}(\zeta_1,\zeta_2,\zeta_3)+\cdots=\hat{G}
\end{array}\right.
\end{eqnarray}
Symmetries of system (\ref{sys2}), which are generated
by elements of $\ring$, can be represented by
evolutionary equations
\begin{eqnarray}
\label{symsymb}
&&u_{\tau}=u\Omega_{1}(\xi_1)+v\Omega_{2}(\zeta_1)+u^2A_{20}(\xi_1,\xi_2)+
uvA_{11}(\xi_1,\zeta_1)+v^2A_{02}(\zeta_1,\zeta_2)+u^3A_{30}(\xi_1,\xi_2,\xi_3)+\\
\nonumber
&&+u^2vA_{21}(\xi_1,\xi_2,\zeta_1)+uv^2A_{12}(\xi_1,\zeta_1,\zeta_2)+v^3A_{03}(\zeta_1,\zeta_2,\zeta_3)+\cdots
\end{eqnarray}
where
$\Omega_{1}(\xi_1),\Omega_2(\zeta_1),A_{ij} $ are polynomials.

It follows from (\ref{frechet}) that the Fr\'echet derivative ${\bf
a}_*$ in the symbolic representation has the form
\begin{equation}\label{astar}
\hat{{\bf a}}_*=\left(\begin{array}{cc}
0& \eta\\
\hat{G}_{*,u}& \hat{G}_{*,v} \end{array}\right)
\end{equation}
where
$$
\hat{G}_{*,u}=\omega_1(\eta)+2ua_{20}(\xi_1,\eta)+va_{11}(\eta,\zeta_1)+3u^2a_{30}(\xi_1,\xi_2,\eta)+
2uva_{21}(\xi_1,\eta,\zeta_1)+
v^2a_{12}(\eta,\zeta_1,\zeta_2)+\cdots
$$
$$
\hat{G}_{*,v}=\omega_2(\eta)+ua_{11}(\xi_1,\eta)+2va_{02}(\zeta_1,\eta)+u^2a_{21}(\xi_1,\xi_2,\eta)+
2uva_{12}(\xi_1,\zeta_1,\eta)+
3v^2a_{03}(\zeta_1,\zeta_2,\eta)+\cdots
$$

Let us define a $2\times 2$ matrix $\hat{R}$ whose entries are
formal series
\begin{eqnarray}
\label{rec1}
\hat{R}_{11}&=&\phi_{00}(\eta)+u\phi_{10}(\xi_1,\eta)+\frac{1}{2}v\phi_{01}(\zeta_1,\eta)+
u^2\phi_{20}(\xi_1,\xi_2,\eta)+
uv\phi_{11}(\xi_1,\zeta_1,\eta)+v^2\phi_{02}(\zeta_1,\zeta_2,\eta)+\cdots\\ \label{rec2}
\hat{R}_{12}&=&\psi_{00}(\eta)+\frac{1}{2}u\psi_{10}(\xi_1,\eta)+v\psi_{01}(\zeta_1,\eta)+
u^2\psi_{20}(\xi_1,\xi_2,\eta)+
uv\psi_{11}(\xi_1,\zeta_1,\eta)+v^2\psi_{02}(\zeta_1,\zeta_2,\eta)+\cdots\, ,
\end{eqnarray}
and (compare with (\ref{R21R22}))
\begin{equation}\label{rhat2122}
\hat{R}_{21}=\eta^{-1}\circ(\hat{R}_{11,t}+\hat{R}_{12} \circ
\hat{G}_{*,u}),\quad
\hat{R}_{22}=\eta^{-1}\circ(\hat{R}_{12,t}+\hat{R}_{12} \circ
\hat{G}_{*,v}+\hat{R}_{11}\circ  \eta)
\end{equation}
\begin{Def}
We shall call matrix $\hat{R}$ a formal recursion operator of system (\ref{syssym}) if it satisfies equation
\begin{equation}
\label{receq}
\hat{R}_t=[\hat{{\bf a}}_*,\hat{R}]
\end{equation}
and all the coefficients $\phi_{ij},\psi_{ij},\,i,j=0,1,2,\ldots$ of the formal series $\hat{R}_{11}$
and $\hat{R}_{12}$ are local.
\end{Def}

Any recursion operator in symbolic representation satisfies the above Definition.
The category of formal series is less restrictive than the category of
pseudo-differential operators, since the action of formal series on
the ring $\ring$ is not defined
and there is no issue of convergence.
Actually we always consider a few first terms of $R_{11}$ and $R_{12}$.
Therefore it is convenient to introduce a notion of an approximate formal recursion operator.

\begin{Def}
We call matrix $\hat{R}$ an approximate formal recursion operator of degree $n$ if it satisfies
\[
\hat{R}_t-[\hat{{\bf a}}_*,\hat{R}]=o(\hat{\ring}^n[\eta])
\]
and all the coefficients up to degree $n$  of  formal series $R_{11}$ and $R_{12}$ are local.
\end{Def}

\begin{The}\label{nc}
 Suppose that system (\ref{syssym}) possess infinitely many
approximate symmetries (\ref{symsymb}) of degree $m$
\[
u_{\tau_i}=u\Omega_{1,i}(\xi_1)+v\Omega_{2,i}(\zeta_1)+o(\ring^1)\,  .
\]
\begin{itemize}
\item If $\ 0<\deg(\Omega_{1,1})<\deg(\Omega_{1,2})<\deg(\Omega_{1,3})<\ldots\ $,
then system (\ref{syssym}) possess an approximate formal recursion
operators of degree $m-1$ with
$\phi_{00}(\eta)=\eta,\quad\psi_{00}(\eta)=0$.

\item If $\ 0<\deg(\Omega_{2,1})<\deg(\Omega_{2,2})<\deg(\Omega_{2,3})<\ldots\ $,
then system (\ref{syssym}) possess an approximate formal recursion
operators of degree $m-1$ with
$\phi_{00}(\eta)=0,\quad\psi_{00}(\eta)=\eta$.
\end{itemize}
\end{The}

The proof of the Theorem is straightforward and very similar to the
proof of the existence of a formal recursion operator given in
\cite{mn}.

In symbolic representation equation (\ref{receq}) can be solved and
all the coefficients $\phi_{ij},\psi_{ij},\,i,j=0,1,2,3,\ldots$ can be
expressed in terms of the right-hand side of system (\ref{syssym}),
i.e. $\omega_i,a_{ij}$. A substitution of (\ref{astar}),(\ref{rec1}),(\ref{rec2}),(\ref{rhat2122})
in (\ref{receq}) leads to a system of linear algebraic
equations on the coefficient functions of the formal recursion
operator. There are no restrictions on the choice of $\phi_{00}
(\eta), \psi_{00} (\eta)$ and they can be chosen arbitrary (for
example as in the above Theorem:
$\phi_{00}(\eta)=\eta,\quad\psi_{00}(\eta)=0$ or
$\phi_{00}(\eta)=0,\quad \psi_{00}(\eta)=\eta$). To determine
coefficients   $\phi_{ij} (\xi,\eta), \psi_{ij} (\xi,\eta)$ when
$i+j=1$, we have to solve a system of four linear
equations.

In what follows we will study homogeneous equations (\ref{eq1}) of
even order $p=2n$, which can be written in the form (\ref{sys2}). Thus in (\ref{syssym}) we assume
\[ \omega_1(\xi)=\alpha\xi^{2n-1}\, ,\qquad \omega_2(\xi)=\beta\xi^{n}\, ,\]
where $\alpha,\beta $ are constant parameters. Without loss of
generality we can choose
\[
\alpha=\frac{\mu^2-\beta^2}{4}.
\]
If $\mu\ne 0,\pm\beta$ then the solution of equations for
$\phi_{10} (\xi,\eta), \psi_{10} (\xi,\eta), \phi_{01} (\xi,\eta),
\psi_{01} (\xi,\eta)$ can be written in the form
\begin{eqnarray*}
\phi_{10}(\xi,\eta)&=&\frac{1}{16\mu^2}\bigg[(\beta-\mu)^2f_1(\mu,\xi,\eta)+(\beta+\mu)^2f_1(-\mu,\xi,\eta)+
(\mu^2-\beta^2)(f_2(\mu,\xi,\eta)+f_2(-\mu,\xi,\eta))\bigg],\\
\psi_{10}(\xi,\eta)&=&\frac{1}{4\mu^2\eta^{n-1}}\bigg[(\beta-\mu)f_1(\mu,\xi,\eta)+(\beta+\mu)f_1(-\mu,\xi,\eta)+
(\mu-\beta)f_2(\mu,\xi,\eta)-(\mu+\beta)f_2(-\mu,\xi,\eta))\bigg],\\
\phi_{01}(\xi,\eta)&=&\frac{1}{4\mu^2\xi^{n-1}}\bigg[(\beta-\mu)f_1(\mu,\xi,\eta)+(\beta+\mu)f_1(-\mu,\xi,\eta)-
(\mu+\beta)f_2(\mu,\xi,\eta)+(\mu-\beta)f_2(-\mu,\xi,\eta))\bigg],\\
\psi_{01}(\xi,\eta)&=&\frac{1}{4\mu^2\xi^{n-1}\eta^{n-1}}\bigg[f_1(\mu,\xi,\eta)+f_1(-\mu,\xi,\eta)-
f_2(\mu,\xi,\eta)-f_2(-\mu,\xi,\eta))\bigg],
\end{eqnarray*}
where
\[
f_1(\mu,\xi,\eta )=\frac{M_1(\mu,\xi,\eta )r_1(\mu,\xi,\eta
)}{g_1(\mu,\xi,\eta )g_4(-\mu,\xi,\eta )},\qquad \nonumber
f_2(\mu,\xi,\eta )=\frac{M_2(\mu,\xi,\eta )r_2(\mu,\xi,\eta
)}{g_2(\mu,\xi,\eta )g_3(-\mu,\xi,\eta )}
\]
and
\begin{eqnarray}
\nonumber M_1(\mu,\xi,\eta )&=&-(\xi+\eta
)\bigg[2(\phi_{00}(\xi+\eta )-\phi_1(\eta ))+(\beta+\mu)((\eta
^n+\xi^n)\psi_{00}(\xi+\eta )
-\eta ^n\psi_{00}(\eta ))\bigg],\\
\nonumber M_2(\mu,\xi,\eta )&=&-(\xi+\eta
)\bigg[2(\phi_{00}(\xi+\eta )-\phi_{00}(\eta ))+(\beta-\mu)\eta
^n(\psi_{00}(\xi+\eta )-\psi_{00}(\eta ))+(\beta+\mu)
\xi^n\psi_{00}(\xi+\eta )\bigg]
\end{eqnarray}
\begin{eqnarray}
\nonumber
g_1(\mu,\xi,\eta )&=&(\beta-\mu)(\xi+\eta )^n-(\beta+\mu)(\xi^n+\eta ^n),\\
\nonumber
g_2(\mu,\xi,\eta )&=&(\beta-\mu)(\xi+\eta )^n-(\beta+\mu)\xi^n-(\beta-\mu)\eta ^n,\\
\nonumber
g_3(\mu,\xi,\eta )&=&(\beta-\mu)(\xi+\eta )^n-(\beta-\mu)\xi^n-(\beta+\mu)\eta ^n,\\
\nonumber g_4(\mu,\xi,\eta )&=&(\beta-\mu)((\xi+\eta )^n-\xi^n-\eta
^n)
\end{eqnarray}
\begin{eqnarray}
\nonumber r_{1}(\mu,\xi,\eta )&=&4a_{20}(\xi,\eta )+(\beta+\mu)\eta
^{n-1}a_{11}(\xi,\eta
)+(\beta+\mu)\xi^{n-1}a_{11}(\eta,\xi)+(\beta+\mu)^2\xi^{n-1} \eta
^{n-1}a_{02}(\xi,\eta ),\\
\nonumber r_{2}(\mu,\xi,\eta )&=&4a_{20}(\xi,\eta )+(\beta-\mu)\eta
^{n-1}a_{11}(\xi,\eta
)+(\beta+\mu)\xi^{n-1}a_{11}(\eta,\xi)+(\beta^2-\mu^2)\xi^{n-1}\eta
^{n-1}a_{02}(\xi,\eta )
\end{eqnarray}

In a similar way, solving a linear system of 8 equations, one can
find coefficients $\phi_{20},\phi_{11}, \phi_{02},
\psi_{20},\psi_{11}, \psi_{02}$, etc. Despite of a straightforward
way of computation, the explicit expressions for the coefficients
$\phi_{20},\ldots  \psi_{02}$ are quite cumbersome and, therefore we
do not write them down in the paper.

Cases $\alpha =0$ (i.e. $\mu=\pm \beta$) and $4\alpha+\beta^2=0$ (i.e. $\mu=0$)
require a consideration of higher degree terms and
have to be treated separately.



\section{Integrable equations of 4th, 6th and 10th order}

In this section we consider homogeneous partial differential equations of the form
\begin{equation}\label{sys2aa}
u_t=v_1, \, \quad v_t=\alpha u_{2n-1}+\beta v_n+G(u,v,\ldots
u_{2n-2},v_{n-1}),
\end{equation}
with $n=2,3,5$, assuming that the polynomial $G(u,v,\ldots
u_{2n-2},v_{n-1})$ has non-zero quadratic terms and parameters
$\alpha$ and $\beta$ do not vanish simultaneously. We also assume
that the weight $W(u)$ of the variable $u$ is positive (it is
automatically integer, since $G$ has a quadratic part). It follows
from (\ref{sys2aa}) that the weight of the variable $v$ is
$W(v)=W(u)+n-1$.  In the case $\partial G/\partial v=0$,  system
(\ref{sys2aa}) corresponds to a non-evolutionary equation of form
(\ref{eq1}) with of order $p=2n$ (and $q=n$).

We select equations (\ref{sys2aa}) that satisfy necessary
conditions for the existence of higher symmetries, i.e. the
conditions that a first few coefficients of the corresponding formal
recursion operator are local functions and then prove their
integrability (the existence of an infinite hierarchy of higher
symmetries). For all equations found we show that they satisfy
sufficient conditions for integrability, such as they possess
bi-Hamiltonian structures, recursion operators and the Lax
representations or linearising substitutions.

\subsection{The 4th order equations}

It is easy to see that a homogeneous system (\ref{sys2aa}) with $n=2$  is linear if $W(u)> 3$.
In the case $W(u)=3$ the only possibility is $G=\gamma u^2,\ \gamma\ne 0$ which leads to a
non-integrable equation for any choice of $\alpha,\beta$ and $\gamma$. Thus the the weight
$W(u)$ can be equal to $2$ or $1$.

\subsubsection{The case of $W(u)=2$}
The most general nonlinear homogeneous system of equations (\ref{sys2aa}) with $W(u)=2$
(correspondingly $W(v)=3$) is of the form:
\begin{eqnarray}
\label{eq4h2}
\left\{\begin{array}{l}
u_t=v_1\\
v_t=\alpha u_3+\beta v_2+c_1uu_1+c_2uv,
\end{array}\right.
\end{eqnarray}
where $c_1,c_2$ are arbitrary constants and at least one of them is not zero.

Without loss of generality we need to consider the following three
types of system (\ref{eq4h2}):
\begin{eqnarray}
\label{eq4h2_1} &&\left\{\begin{array}{l}
u_t=v_1\\
v_t=\frac{\mu^2-\beta^2}{4} u_3+\beta v_2+c_1uu_1+c_2uv,\quad
\mu\notin \{0,\pm\beta\},\,\,\,\mu,\beta\in {\bbbc},
\end{array}\right.\\
\label{eq4h2_2} &&\left\{\begin{array}{l}
u_t=v_1\\
v_t=v_2+c_1uu_1+c_2uv
\end{array}\right.\\
\label{eq4h2_3} &&\left\{\begin{array}{l}
u_t=v_1\\
v_t=-\frac{1}{4}u_3+v_2+c_1uu_1+c_2uv
\end{array}\right.
\end{eqnarray}
The first system represents the generic case $\alpha\notin \{0, -\frac{\beta^2}
{4}\}$, while the other two represent the degenerate
cases.

\begin{Pro} System (\ref{eq4h2_1}) possess two formal recursion operators with
$\phi_{00}(\eta)=\eta,\,\psi_{00}(\eta)=0$
and $\phi_{00}(\eta)=0,\,\psi_{00}(\eta)=\eta$ if and only if $\beta=c_2=0$.
By re-scalings it can be put in the form
\begin{eqnarray}
\label{eq4h2a}
\left\{\begin{array}{l}
u_t=v_1\\
v_t=u_3+2uu_1
\end{array}\right.
\end{eqnarray}
Systems (\ref{eq4h2_2}) and (\ref{eq4h2_3}) are not integrable, they do not possess a formal
recursion operator with $\phi_{00}=0,\,\psi_{00}=\eta$ unless
$c_1=c_2=0$.
\end{Pro}
System (\ref{eq4h2a}) represents the Boussinesq equation (\ref{bouss1}),
which is known to be integrable. Its Lax representation can be found in \cite{zakharov}.
Recursion, Hamiltonian and symplectic operators for the Boussinesq equation were given in Section \ref{RHS}. In the Proposition 1 and below by re-scalings we mean an invertible change
of variable of the form:
\[
u\to \alpha_1\,u,\quad  v\to \alpha_2\,v,\quad x\to \alpha_3\,x,\quad t\to \alpha_4\,t,\qquad \alpha_i\in\bbbc\, .\]

\subsubsection{The case of $W(u)=1$}
The most general homogeneous system of equations (\ref{sys2aa}) corresponding to the case $W(u)=1$ and $n=2$ is:
\begin{eqnarray}
\label{eq4h1}
\left\{\begin{array}{l}
u_t=v_1\\
v_t=\alpha u_3+\beta v_2+c_1uu_2+c_2u_1^2+c_3u_1v+c_4uv_1+c_5v^2+c_6u^2u_1+c_7u^2v+c_8u^4
\end{array}\right.
\end{eqnarray}
where $c_i,\,i=1\ldots 8$ are arbitrary constants and we assume that
at least one of the coefficients $c_1,\ldots, c_5$ is not zero.
Without loss of generality we consider the following three types of
the system (\ref{eq4h1}):

\begin{eqnarray}
\label{eq4h1_1} \left\{\begin{array}{l}
u_t=v_1\\
v_t=\frac{\mu^2-\beta^2}{4} u_3+\beta
v_2+c_1uu_2+c_2u_1^2+c_3u_1v+c_4uv_1+c_5v^2+c_6u^2u_1+c_7u^2v+c_8u^4,
\end{array}\right.
\end{eqnarray}
where $\mu\notin \{0,\pm\beta\},\,\,\,\mu,\beta\in {\bbbc}$ and
\begin{eqnarray}
\label{eq4h1_2} &&\left\{\begin{array}{l}
u_t=v_1\\
v_t=v_2+c_1uu_2+c_2u_1^2+c_3u_1v+c_4uv_1+c_5v^2+c_6u^2u_1+c_7u^2v+c_8u^4
\end{array}\right.\\
\label{eq4h1_3} &&\left\{\begin{array}{l}
u_t=v_1\\
v_t=-\frac{1}{4}u_3+v_2+c_1uu_2+c_2u_1^2+c_3u_1v+c_4uv_1+c_5v^2+c_6u^2u_1+c_7u^2v+c_8u^4
\end{array}\right.
\end{eqnarray}

\begin{Pro}
System (\ref{eq4h1_1}) possess two  formal recursion operators  with
$\phi_{00}(\eta)=\eta,\,\psi_{00}(\eta)=0$ and
$\phi_{00}(\eta)=0,\,\psi_{00}(\eta)=\eta$ if and only if (up to
re-scalings) it is
one of the list
\begin{eqnarray}
&&\left\{\begin{array}{l}
u_t=v_1\\
v_t=u_3+u_1^2
\end{array}\right.
 \label{eq4h1a}\\
&&\left\{\begin{array}{l}
u_t=v_1\\
v_t=u_3+2u_1v+2u^2u_1
\end{array}\right.
\label{eq4h1b}\\
&&\left\{\begin{array}{l}
u_t=v_1\\
v_t=u_3+2u_1v+4uv_1-6u^2u_1
\end{array}\right.
\label{eq4h1c}\\
&&\left\{\begin{array}{l}
u_t=v_1\\
v_t=u_3+4uu_2+3u_1^2-v^2+6u^2u_1+u^4
\end{array}\right.
\label{eq4h1d}\\
&&\left\{\begin{array}{l}
u_t=v_1\\
v_t=\alpha u_3+v_2+4 \alpha u u_2+3 \alpha u_1^2+u_1 v +2 u
v_1-v^2+6 \alpha u^2 u_1+u^2 v+\alpha u^4,\,\,\alpha\ne -\frac{1}{4}
\end{array}\right. \label{eq4h1e}
\end{eqnarray}
System (\ref{eq4h1_2}) possess a formal recursion operator  with
$\phi_{00}(\eta)=0,\,\psi_{00}(\eta)=\eta$ if and only if (up to
re-scalings) it is
one of the list
\begin{eqnarray}
&&\left\{\begin{array}{l}
u_t=v_1\\
v_t=v_2+2 u v_1
\end{array}\right.
\label{eq4h1f}\\
&&\left\{\begin{array}{l}
u_t=v_1\\
v_t=v_2-u_1^2+2 u_1 v-v^2
\end{array}\right.
\label{eq4h1g}\\
&&\left\{\begin{array}{l}
u_t=v_1\\
v_t=v_2-2 u u_2-2 u_1^2+2 u_1 v+6 u v_1-12 u^2 u_1
\end{array}\right.\label{eq4h1h}
\end{eqnarray}
System (\ref{eq4h1_3}) possess two formal recursion operators  with
$\phi_{00}(\eta)=\eta,\,\psi_{00}(\eta)=0$ and
$\phi_{00}(\eta)=0,\,\psi_{00}(\eta)=\eta$  if and only if (up to
re-scalings) it is
\begin{eqnarray}
\left\{\begin{array}{l}
u_t=v_1\\
v_t=-\frac{1}{4}u_3+v_2-u u_2-\frac{3}{4}u_1^2+u_1 v +2 u
v_1-v^2-\frac{3}{2}u^2 u_1+u^2 v-\frac{1}{4} u^4.
\end{array}\right. \label{eq4h1ep}
\end{eqnarray}

\end{Pro}

Only equations (\ref{eq4h1a}) and (\ref{eq4h1f}) from this list
represent second order equations of the form  (\ref{eq1}) in the
standard set of dynamical variables, since their right hand side do
not depend on the variable $v$ explicitly. Equation (\ref{eq4h1a})
is a potential version of the Boussinesq equation, cf. (\ref{potbouss}).
Indeed, if we introduce variables $U_1=u, V_1=v$ then in terms of $U,V$
equation (\ref{eq4h2a}) takes form (\ref{eq4h1a}). Its symmetries,
conservation laws, recursion and multi-Hamiltonian structures can be
easily recovered  from the theory of the Boussinesq equation. For
example, the recursion operator of the potential Boussinesq equation
(\ref{eq4h1a}) is $D_x^{-1}\circ R \circ D_x$ where $R$ is the
recursion operator (\ref{recboussR}) in which $u_k,v_k$ is replaced
by $u_{k+1},v_{k+1}$ and therefore the corresponding restricted
recursion operator for equation (\ref{eq4h1a}) is of the form
\[
\Re=D_x^{-1}\circ \left( 3v_1D_x+2v_2+(4D_x^{2}+2u_1+u_2D_x^{-1})\circ D_t\right)\, .
\]

Equations (\ref{eq4h1b}) and (\ref{eq4h1c}) are known to be integrable.
Corresponding Lax representations and references can be found in \cite{Cl}.

Equations (\ref{eq4h1d}) and  (\ref{eq4h1e})  can be mapped into linear equations
$$ w_{tt}=w_4\, ,\qquad w_{tt}=\alpha w_4+w_{2t} $$
respectively by the Cole-Hopf transformation $u=(\log w)_x$. In
these cases the restricted recursion operator is the same as for the
Burgers equation
\begin{equation}\label{colehopfr}
\Re=D_x+u+u_x D_x^{-1}\, ,
\end{equation}
which generates higher symmetries from the seeds $u_1$ and $u_t$.

Equation (\ref{eq4h1f}) can be reduced to the Burgers equation with
time independent forcing
$$ u_t=u_2+2 uu_1+w_1 \, ,\qquad w_t=0$$
by invertible transformation $v=w+u_1+u^2$. The latter can be
linearised by the Cole-Hopf transformation.

Equation (\ref{eq4h1g}) can be reduced to the system
$$u_t=u_2+w_1\, ,\qquad w_t=-w^2$$
 by a simple invertible the change of the variable $v=w+u_1$. System (\ref{eq4h1g}) provides
an example of an equation that possesses neither higher symmetries
nor a recursion operator. However, a formal recursion operator does
exist and therefore it is in the list of the Proposition. Its
integration can be reduced to the integration of a linear
nonhomogeneous heat equation with a source term of a special form.

System (\ref{eq4h1h}) possesses an infinite hierarchy of symmetries
of all orders generated by a recursion operator
\[ \Re=-2u+2 u_1 D_x^{-1}+D_t D_x^{-1}\, \]
starting from the seed $u_1$.

By a simple shift of the variable $v=w+u_1+2 u^2$,
system (\ref{eq4h1h}) can be transformed in the form
\begin{equation}\label{eq4h1hsys}
 u_t=u_2+4u u_1 +w_1\, ,\qquad w_t=2D_x (uw)\, .
\end{equation}
This system can be linearised by a reciprocal and then point transformations
(we would like to thank A. Hone who helps us to find an explicit form of the
linearising transformation).

Equation (\ref{eq4h1ep}) is a particular case of (\ref{eq4h1e}) corresponding to the exceptional case $\alpha=-\frac{1}{4}$.

\subsection{The 6th order equations}
Homogeneous 6th order ($n=3$) equations (\ref{sys2aa}) with non-zero
quadratic terms correspond to $W(u)\le 5$. We restrict ourself with
the case $W(u)>0$.  The weights $3,4,5$ do not lead to integrable
equations:

\begin{Pro} For weights $W(u)=3, 4, 5$ there are no equations  possessing a formal recursion operator.
\end{Pro}

\subsubsection{The case of $W(u)=2$}
The most general homogeneous system (\ref{sys2aa}) corresponding to $W(u)=2$ can be written as
\begin{eqnarray}
&&\left\{\begin{array}{l}
u_t=v_1\\
v_t=\alpha u_5+\beta v_3+D_x[c_1uu_2+c_2u_1^2+c_5u^3]+c_3uv_1+c_4vu_1,
\end{array}\right.
 \label{eq6h2}
\end{eqnarray}
where $\alpha, \beta ,c_i,i=1,\ldots,4$ are arbitrary
constants and we assume that at least one of $c_1,\ldots, c_4$ is not zero.

Without loss of generality we consider the following three types of
the above system:
\begin{eqnarray}
&&\left\{\begin{array}{l}
u_t=v_1\\
v_t=\frac{\mu^2-\beta^2}{4} u_5+\beta
v_3+D_x[c_1uu_2+c_2u_1^2+c_5u^3]+c_3uv_1+c_4vu_1,\quad \mu\notin
\{0,\pm\beta\},\,\,\,\mu,\beta\in {\bbbc}
\end{array}\right.
 \label{eq6h2_1}\\
&&\left\{\begin{array}{l}
u_t=v_1\\
v_t=v_3+D_x[c_1uu_2+c_2u_1^2+c_5u^3]+c_3uv_1+c_4vu_1
\end{array}\right.
 \label{eq6h2_2}\\
&&\left\{\begin{array}{l}
u_t=v_1\\
v_t=-\frac{1}{4}
u_5+v_3+D_x[c_1uu_2+c_2u_1^2+c_5u^3]+c_3uv_1+c_4vu_1
\end{array}\right.
 \label{eq6h2_3}
 \end{eqnarray}

\begin{Pro}\label{pro3}
If system (\ref{eq6h2_1}) possess two formal recursion operators
with $\phi_{00}(\eta)=\eta,\,\psi_{00}(\eta)=0$ and
$\phi_{00}(\eta)=0,\,\psi_{00}(\eta)=\eta$ then, up to re-scalings, it is one of
the list
\begin{eqnarray}
&&\left\{\begin{array}{l}\label{eq6h2a}
u_t=v_1\\
v_t=2u_5+v_3+D_x[2uu_2+u_1^2+\frac{4}{27}u^3]
\end{array}\right.
\\
&&\left\{\begin{array}{l}\label{eq6h2b}
u_t=v_1\\
v_t=\frac{1}{5}u_5+v_3+D_x[uu_2+uv+\frac{1}{3}u^3]
\end{array}\right.
\\
&&\left\{\begin{array}{l} \label{eq6h2c}
u_t=v_1\\
v_t=\frac{1}{5}u_5+v_3+D_x[2uu_2+\frac{3}{2}u_1^2+2uv+\frac{4}{3}u^3]
\end{array}\right.
\end{eqnarray}
If system (\ref{eq6h2_2}) possess a formal recursion operator with
$\phi_{00}(\eta)=0,\,\psi_{00}(\eta)=\eta$ then, up to re-scalings, it is one of
the list
\begin{eqnarray}
&&\left\{\begin{array}{l} \label{eq6h2d}
u_t=v_1\\
v_t=v_3+u v_1+u_1 v
\end{array}\right.
\\
&&\left\{\begin{array}{l} \label{eq6h2e}
u_t=v_1\\
v_t=v_3+2 u u_3+4 u_1 u_2-4 u_1 v-8 u v_1-24 u^2 u_1
\end{array}\right.
\end{eqnarray}
System (\ref{eq6h2_3}) does not possess a formal recursion operator
for any non-trivial choice of
$c_i$.
\end{Pro}

Equation (\ref{eq6h2a}) can be rewritten in the form (\ref{eq1}):
\[
u_{tt}=2u_6+u_{3,t}+D_x^2\left(2uu_2+u_1^2+\frac{4}{27}u^3\right)
\]
It has the following restricted recursion operator
\begin{eqnarray*}
\Re&=&D_x^4+\frac{5}{2}D_x D_t +\frac{1}{3}u_t D_x^{-1}
+\frac{5}{9}u D_x^2+\frac{2}{9}u D_t D_x^{-1}\\
&&+\frac{13}{9}u_1 D_x +\frac{7}{9}u_2 +\frac{4}{9} v
+\frac{1}{9}u_1 D_t D_x^{-2}.
\end{eqnarray*}
Applying powers of $\Re$ to the seed symmetries $u_1$
and $u_t$, one can build up two infinite hierarchies of commuting
symmetries generated by $S_{4n+1}=\Re^n(u_1)$ and
$S_{4n+3}=\Re^n(u_t)$ respectively. There are no even order
symmetries for equation (\ref{eq6h2a}).
Its bi-Hamiltonian structure is given by
a Hamiltonian operator
\[ \cH_{11}=\frac{5}{2}D_x^3+\frac{1}{9} (u D_x +D_x u)\]
\[ \cH_{12}=\frac{7}{2} D_x^5+\frac{4}{9} v D_x +\frac{1}{3} v_1
+\frac{7}{9} u D_x^3+\frac{14}{9} u_1 D_x^2 +\frac{7}{9} u_2 D_x\]
and a symplectic operator
\[
\cJ_{12}=D_x^{-1}, \quad \cJ_{22}=0 \ .
\]

A simple change of the variable $v=w-u_2-u^2/3$ brings system  (\ref{eq6h2a}) in the form
\[ u_t=-u_3-\frac{2}{3}uu_1+w_1\, ,\qquad w_t=2w_3+\frac{2}{3}uw_1\, . \]
Thus system (\ref{eq6h2a}) has a reduction ($w=0$, i.e.
$v=-u_2-u^2/3$) to the Korteweg de Vries equation.

This system has the following  Lax representation:
\begin{eqnarray}
\nonumber
L&=&D_x^4+\frac{2}{9}uD_x^2+\frac{2}{9}u_1D_x+\frac{5}{54}u_2
+\frac{1}{162}u^2-\frac{1}{54}v,\\
\nonumber A&=&4D_x^3+\frac{2}{3}u D_x+\frac{1}{3}u_1, .
\end{eqnarray}
The above mentioned reduction of the system corresponds to the case when
$L$ is the square of a second order operator (i.e. the square of the Lax operator for the KdV equation):
\[ L=(D_x^2+\frac{1}{9}u)^2\, .
\]

System (\ref{eq6h2a}) and its Lax representation can be transformed to one of the cases
studied in \cite{DS1}.

Recursion operators and bi-Hamiltonian structure as well as the Lax representations for the
potential versions of equations (\ref{eq6h2b}) and
(\ref{eq6h2c}) will be given in the next section.

Equation (\ref{eq6h2d}) is known to be integrable, its Lax representation can be found in \cite{DS2}.
Its bi-Hamiltonian structure is given by a Hamiltonian operator
\[ \cH_{11}= 5 D_x^3 +2 u D_x +u_1\]
\[ \cH_{12}=\frac{9}{2} (D_x^5+ u D_x^3+2 u_1 D_x^2+ u_2 D_x)
+4 v D_x +3 v_1\]
and a symplectic operator
\[ \cJ_{21}= D_x+ u D_x^{-1},\quad  \cJ_{22}=2 D_x^{-1}. \]
Its restricted recursion operator is of order 6. The hierarchies of
symmetries can be generated from the seeds
$S_1=u_1$, $S_3=u_t$ and $S_5=u_5-10 v_3+D_x(5 u u_2-10 uv+\frac{5}{3} u^3)$.

Equation (\ref{eq6h2e}) has the following restricted recursion operator
\[ \Re =\frac{1}{2} D_x D_t+ u D_x^2+ u_1 D_x +u_2-10 u^2-2 v-2 u D_x^{-1}D_t
-v_1D_x^{-1}-4u_1D_x^{-1} u-u_1 D_x^{-2} D_t \, ,\]
which is of weight $4$. There are  two infinite hierarchies of
symmetries $S_{4k+1}=\Re^k (u_1)$ and $S_{4k+3}=\Re^k(v_1)$
and no symmetries of even order.

Its bi-Hamiltonian structure is given by a Hamiltonian operator
\[ \cH_{11}= -\frac{1}{2} D_x^3 +2 u D_x +u_1\]
\[ \cH_{12}=-\frac{1}{2} D_x^5+5 u D_x^3+10 u_1 D_x^2+7 u_2 D_x+2 u_3
-6 u^2 D_x -12 u u_1+2 v D_x +v_1\]
and a symplectic operator
\[
\cJ_{12}=D_x^{-1}, \quad \cJ_{22}=0 \ .
\]

If we introduce a new variable $v=w + u_2 -3 u^2$, equation
(\ref{eq6h2e}) can be rewritten in the form\footnote{When our paper
was completed we discovered that integrable system (\ref{kdvdef})
has recently been studied by A.B.Shabat \cite{shabat}, he also has
found a Lax representation for this system. }
\begin{equation}\label{kdvdef}
\left\{\begin{array}{l}
u_t=u_3-6 u u_1+w_1\\
w_t=-2 u w_1 -4 u_1 w
\end{array}\right. .
\end{equation}
It admits an obvious reduction ($w=0$) to the KdV equation
$u_t=v_1=u_3-6 u u_1$.

\subsubsection{The case of $W(u)=1$}

Homogeneous systems of equations (\ref{sys2aa}) with $W(u)=1$ can be written in the form:
\begin{eqnarray}
\left\{\begin{array}{l}\label{eq6h1}
u_t=v_1\\
v_t=\alpha u_5+\beta v_3+c_1u_2^2+c_2u_1u_3+c_3uu_4+c_4u_2v+c_5u_1v_1+c_6uv_2+c_7v^2+c_8u_1^3+\\
\quad\quad +c_9uu_1u_2 +c_{10}u^2u_3+c_{11}u^2v_1+c_{12}uu_1v+
+c_{13} u^2u_1^2+ c_{14} u^3u_2+c_{15} u^3v+c_{16} u^4u_1+c_{17} u^6
\end{array}\right. ,
\end{eqnarray}
where $\alpha, \beta \in{\bbbc}$, all $c_i,i=1,\ldots,17$ are
arbitrary constants and at least one of $c_1,\ldots c_7$ is not
zero.

We need to consider the following cases of the system (\ref{eq6h1}):
\begin{eqnarray}
\left\{\begin{array}{l}\label{eq6h1_1}
u_t=v_1\\
v_t=\frac{\mu^2-\beta^2}{4} u_5+\beta v_3+c_1u_2^2+c_2u_1u_3+c_3uu_4+c_4u_2v+c_5u_1v_1+c_6uv_2+c_7v^2+c_8u_1^3+\\
\quad\quad +c_9uu_1u_2 +c_{10}u^2u_3+c_{11}u^2v_1+c_{12}uu_1v+
+c_{13} u^2u_1^2+ c_{14} u^3u_2+c_{15} u^3v+c_{16} u^4u_1+c_{17}
u^6
\end{array}\right.
\end{eqnarray}
where $\mu\notin \{0,\pm\beta\},\,\,\mu,\beta\in {\bbbc}$, and
\begin{eqnarray}
&&\left\{\begin{array}{l}\label{eq6h1_2}
u_t=v_1\\
v_t=v_3+c_1u_2^2+c_2u_1u_3+c_3uu_4+c_4u_2v+c_5u_1v_1+c_6uv_2+c_7v^2+c_8u_1^3+\\
\quad\quad +c_9uu_1u_2 +c_{10}u^2u_3+c_{11}u^2v_1+c_{12}uu_1v+
+c_{13} u^2u_1^2+ c_{14} u^3u_2+c_{15} u^3v+c_{16} u^4u_1+c_{17} u^6
\end{array}\right.\\
&& \left\{\begin{array}{l}\label{eq6h1_3}
u_t=v_1\\
v_t=-\frac{1}{4}u_5+v_3+c_1u_2^2+c_2u_1u_3+c_3uu_4+c_4u_2v+c_5u_1v_1+c_6uv_2+c_7v^2+c_8u_1^3+\\
\quad\quad +c_9uu_1u_2 +c_{10}u^2u_3+c_{11}u^2v_1+c_{12}uu_1v+
+c_{13} u^2u_1^2+ c_{14} u^3u_2+c_{15} u^3v+c_{16} u^4u_1+c_{17} u^6
\end{array}\right.
\end{eqnarray}

\begin{Pro}
If system (\ref{eq6h1_1}) possess two  formal recursion operators
with $\phi_{00}(\eta)=\eta,\,\psi_{00}(\eta)=0,$  and
$\phi_{00}(\eta)=0,\,\psi_{00}(\eta)=\eta,$ up to re-scalings, it is one of the
equations in the following list
\begin{eqnarray}
&&\left\lbrace \begin{array}{l}\label{eq6h1c} u_t=v_1\\
v_t=2u_5+v_3+u_2^2+2u_1u_3+\frac{4}{27}u_1^3\end{array}
\right. \\
&&\left\lbrace \begin{array}{l}
\label{eq6h1a} u_t=v_1\\
v_t=\frac{1}{5}u_5+v_3+u_1u_3+u_1v_1+\frac{1}{3}u_1^3\end{array}
\right. \\
&&\left\lbrace \begin{array}{l}\label{eq6h1b} u_t=v_1\\
v_t=\frac{1}{5}u_5+v_3+2u_1u_3+\frac{3}{2}u_2^2+2u_1v_1+\frac{4}{3}u_1^3\end{array}
\right.\\
&&\left\lbrace \begin{array}{l}\label{eq6h1e} u_t=v_1\\
v_t=\alpha u_5+v_3+10\alpha u_2^2+15\alpha u_1u_3+6\alpha
uu_4+vu_2+3u_1v_1+3uv_2-v^2+ 15\alpha u_1^3+15\alpha u^2u_3+\\
\phantom{v_t=} +60\alpha uu_1u_2+3uu_1v+3u^2v_1+45\alpha
u^2u_1^2+20\alpha u^3u_2+u^3v+15\alpha u^4u_1+\alpha
u^6,\,\,\alpha\ne-\frac{1}{4}\end{array}
\right. \\
&&\left\lbrace \begin{array}{l}\label{eq6h1f} u_t=v_1\\
v_t=u_5+6 u u_4+15 u_1 u_3+10 u_1^2-v^2+15 u^2 u_3+15 u_1^3 +60 u
u_1 u_2+45 u^2 u_1^2+\\ \phantom{v_t=}+20 u^3 u_2+15 u^4
u_1+u^6\end{array} \right.
\end{eqnarray}
If system (\ref{eq6h1_2}) possess a  formal recursion operator with
$\phi_{00}(\eta)=0,\,\psi_{00}(\eta)=\eta,$ up to re-scalings, it is one of the list
\begin{eqnarray}
&&\left\lbrace \begin{array}{l}\label{eq6h1d} u_t=v_1\\
v_t=v_3+u_1 v_1\end{array}
\right. \\
&&\left\lbrace \begin{array}{l}\label{eq6h1g} u_t=v_1\\
v_t=v_3+3 u_1 v_1+3 u v_2+3 u^2 v_1\end{array}
\right. \\
&&\left\lbrace \begin{array}{l}\label{eq6h1h} u_t=v_1\\
v_t=v_3-u_2^2+2 u_2 v-v^2\end{array}\right.
\end{eqnarray}
If system (\ref{eq6h1_3}) possess two  formal recursion operators
with $\phi_{00}(\eta)=\eta,\,\psi_{00}(\eta)=0,$  and
$\phi_{00}(\eta)=0,\,\psi_{00}(\eta)=\eta,$ up to re-scalings, then it is
\begin{eqnarray}
\left\lbrace \begin{array}{l}\label{eq6h1ep} u_t=v_1\\
v_t=-\frac{1}{4} u_5+v_3-\frac{5}{2} u_2^2-\frac{15}{4}
u_1u_3-\frac{3}{2}
uu_4+vu_2+3u_1v_1+3uv_2-v^2- \frac{15}{4} u_1^3-\frac{15}{4} u^2u_3+\\
\phantom{v_t=} -15uu_1u_2+3uu_1v+3u^2v_1-\frac{45}{4}
u^2u_1^2-5u^3u_2+u^3v-\frac{15}{4}u^4u_1-\frac{1}{4} u^6,\end{array}
\right.
\end{eqnarray}
\end{Pro}

Equations (\ref{eq6h1c}-\ref{eq6h1b}) and (\ref{eq6h1d})
are ``potential'' versions of  (\ref{eq6h2a}-\ref{eq6h2c}) and (\ref{eq6h2d}).

System (\ref{eq6h1a}) represents in the form of equation  (\ref{eq1}) as
\begin{equation}\label{eq6h1att}
u_{tt}=\frac{1}{5}u_6+u_{3t}+D_x\left(u_1u_3+u_1u_t+\frac{1}{3}u_1^3\right)
\end{equation}

It is a bi--Hamiltonian system with a Hamiltonian
operator
\begin{eqnarray*}
\cH_{11}&=& 6D_x+2u_1D_x^{-1}+2D_x^{-1}\circ u_1\\
\cH_{21}&=&
\frac{33}{5}D_x^3+11u_1D_x+2v_1D_x^{-1}+6D_x^{-1}\circ v_1\, ,
\end{eqnarray*}
and a  symplectic operator is defined by
\begin{eqnarray*}
\cJ_{12}&=&\frac{3}{5}u_1D_x^5+3u_2D_x^4+[5u_3+u_1^2+5v_1]D_x^3+[2u_4+2u_1u_2+5v_2]D_x^2+\\&&+
[-2u_5-2u_1u_3-2u_2^2+4v_3+3u_1v_1]D_x-\frac{4}{5}u_6-2u_2u_3-u_1u_4+v_4+u_2v_1+u_1v_2,\\
\cJ_{22}&=&\frac{27}{5}D_x^5+3u_1D_x^3+D_x^3\circ 3u_1+[-2u_3+\frac{1}{2}u_1^2+v_1]D_x+
D_x\circ [-2u_3+\frac{1}{2}u_1^2+v_1]\, .
\end{eqnarray*}

A restricted recursion operator for equation (\ref{eq6h1att}) can be expressed
in terms of $\cH$ and $\cJ$ (see (\ref{recopHJ})).
This system does not have symmetries of even order
and of order $5\mod 10$.  The hierarchy of symmetries
can be generated by the restricted recursion operator acting on the
seed symmetries $S_{-1} =1$, $S_1=u_1$, $S_3=u_t$ and
\begin{eqnarray*}
S_{7}&=&u_7+7u_{4t}+\frac{14}{3}u_1u_5+\frac{35}{3}u_2u_4
+\frac{35}{9}u_3^2+\frac{70}{9}u_3u_t+\frac{35}{3}u_2u_{1t}+
\frac{35}{3}u_1u_{2t}+\frac{35}{9}u_{t}^2+\\
&&+\frac{35}{9}u_1^2u_3+\frac{35}{3}u_1u_2^2+\frac{35}{9}u_1^2u_t\ .
\end{eqnarray*}
A Lax representation of system (\ref{eq6h1a}) is
\begin{eqnarray}
\nonumber
L&=&D_x^5+\frac{5}{3}u_1D_x^3+\frac{5}{3}u_2D_x^2+\big[\frac{10}{9}u_3
+\frac{5}{9}u_1^2-\frac{5}{9}v_1\big]D_x,\\
\nonumber A&=&D_x^3+ u_1D_x.
\end{eqnarray}

System (\ref{eq6h1b}) represents in the form of equation  (\ref{eq1}) as
$$
u_{tt}=\frac{1}{5}u_6+u_{3,t}+D_x\left(2u_1u_3+\frac{3}{2}u_1^2+2u_1u_t+\frac{4}{3}u_1^3\right)
$$
This is a bi--Hamiltonian system with a Hamiltonian operator
\begin{eqnarray*}
&&{\cal H}_{11}=3 D_x+u_1 D_x^{-1}+D_x^{-1} u_1 \ ,\\
&&{\cal H}_{12}=\frac{12}{5} D_x^3 +4 D_x u_1 +3 u_t D_x^{-1}+D_x^{-1} u_t \ .
\end{eqnarray*}
and a symplectic  operator
\begin{eqnarray*}
\cJ_{12}&=&-\frac{9}{10}D_x^7-\frac{42}{5}u_1D_x^5-\frac{21}{2}u_2D_x^4-[\frac{7}{2}u_3+14u_1^2+14v_1]D_x^3-
[\frac{1}{2}u_4+25u_1u_2+\frac{25}{2}v_2]D_x^2+\\
&&+[\frac{1}{2}u_5+u_1u_3-2u_2^2-\frac{11}{2}v_3-24u_1v_1]D_x+\frac{1}{5}u_6+4u_2u_3+2u_1u_4-v_4-8u_2v_1-8u_1v_2 \ ,\\
\cJ_{22}&=&-\frac{63}{10}D_x^5-\frac{21}{2}(u_1D_x^3+D_x^3 u_1)
+[\frac{19}{2}u_3-4u_1^2-4v_1]D_x+D_x[\frac{19}{2}u_3-4u_1^2-4v_1]\ .
\end{eqnarray*}

A restricted recursion operator can be expressed in terms of $\cH$ and $\cJ$ (see (\ref{recopHJ})).
It is of $10$-th order. The system does not have symmetries
of even order and of order $5\mod 10$.  The hierarchy of symmetries
can be generated by the restricted recursion operator acting on
the seed symmetries
$S_{-1} =1$, $S_1=u_1$, $S_3=u_t$ and
\begin{eqnarray*}
S_{7}&=&u_7+7u_{4t}+\frac{28}{3}u_1u_5+35u_2u_4
+\frac{455}{18}u_3^2+\frac{140}{9}u_3u_t+35u_2u_{1t}+
\frac{70}{3}u_1u_{2t}+\frac{70}{9}u_t^2+\\
&&+\frac{140}{9}u_1^2u_3+35u_1u_2^2+\frac{140}{9}u_1^2 u_t\ .
\end{eqnarray*}
The Lax representation for equation (\ref{eq6h1b}) is :
\begin{eqnarray*}
&&L=D_x^5+\frac{10}{3}u_1D_x^3+5u_2D_x^2+[\frac{35}{9}u_3+\frac{20}{9}u_1^2-\frac{10}{9}v_1]D_x+\frac{10}{9}u_4+\frac{20}{9}u_1u_2-\frac{5}{9}v_2,\\
&&A=D_x^3+2u_1D_x+u_2\, .
\end{eqnarray*}

Equations (\ref{eq6h1e}) and (\ref{eq6h1f})  can be mapped into linear equations
$$ w_{tt}=\alpha w_6+w_{3,t} \, ,\qquad w_{tt}=w_6 $$
respectively by the Cole-Hopf transformation $u=(\log w)_x$. In these cases the restricted
recursion operator is the same as for the Burgers equation, c.f.
(\ref{colehopfr}).

System (\ref{eq6h1g}) represents equation (\ref{eq1}) of the form
\[
u_{tt}=D_t(u_3+3uu_2+3u_1^2+3u^2u_1),
\]
from which follows, that $u_t=u_3+3uu_2+3u_1^2+3u^2u_1+f(x)$, where
$f(x)$ is an arbitrary function. This equation can be linearised by
the Cole-Hopf transformation $u=(\log w)_x$.

Equation (\ref{eq6h1h}) has similar property as system (\ref{eq4h1g}).
It can also be reduced to a triangular system
\[u_t=u_3+w_1,\qquad w_t=-w^2\]
in  variables $u$ and $w=v-u_2$.

System (\ref{eq6h1ep}) is a particular case of (\ref{eq6h1e}) in the exceptional case
$\alpha=-\frac{1}{4}$.

\subsection{10th order equations}
In this section we present three examples of 10th order integrable
non-evolutionary equations together with their bi--Hamiltonian
structures and recursion operators. These equations are new to the
best of our knowledge.

\begin{Pro}
The following systems possess infinite hierarchies of higher
symmetries:
\begin{eqnarray}
\label{ex5} \left\{ \begin{array}{ll}
u_t=v_1\\
v_t=\frac{9}{64}u_9+v_5+D_x\left( 3uu_6+9u_1u_5+\frac{65}{4}u_2u_4
+\frac{35}{4}u_3^2+2u_1v_1+4uv_2+20u^2u_4+80uu_1u_3+\right. \\
\quad\quad
\left. +60uu_2^2+88u_1^2u_2+\frac{256}{5}u^3u_2+\frac{384}{5}u^2u_1^2
+\frac{1024}{125}u^5\right)
\end{array} \right.
\end{eqnarray}
\begin{eqnarray}\label{ex4b}
\left\{\begin{array}{l}
u_t=v_1\\
v_t=-\frac{1}{54}u_{9}+v_5+\frac{5}{6}u_7u_1
+\frac{5}{3}u_6u_2+\frac{5}{2}u_5u_3+\frac{25}{12}u_4^2
-5u_3v_1-\frac{15}{2}u_2v_2-10u_1v_3\\
\quad \quad -\frac{45}{4}u_5u_1^2-\frac{75}{2}u_1u_2u_4
-\frac{75}{4}u_3^2u_1 -\frac{75}{4}u_2^2u_3+\frac{45}{2}u_1^2 v_1+
\frac{225}{4}u_3u_1^3 +\frac{675}{8}u_2^2u_1^2-\frac{405}{16}u_1^5
\end{array}\right.
\end{eqnarray}
\begin{eqnarray}\label{ex4c}
\left\{\begin{array}{l}
u_t=v_1\\
v_t=v_5+2 u_2 u_5+6 u_3 u_4-6
 u_3 v-22 u_2 v_1-30 u_1 v_2-20 u v_3
+96 u u_1 v +96 u^2 v_1
\\ \quad \quad
-2 D_x [ 8 u^2 u_4+32 u u_1 u_3 +13 u_1^2 u_2 +24 u u_2^2]
+120 D_x [4 u^3 u_2 +6 u^2 u_1^2]
-3840 u^4 u_1
\end{array}\right.
\end{eqnarray}
\end{Pro}

System (\ref{ex5}) can be rewritten in the form (\ref{eq1}) as
\begin{eqnarray}
\nonumber
u_{tt}&=&\frac{9}{64}u_{10}+u_{5,t}+D_x^2\bigg(3uu_6+9u_1u_5+\frac{65}{4}u_2u_4
+\frac{35}{4}u_3^2+2u_1u_t+4uu_{1,t}+20u^2u_4+80uu_1u_3+\\
\nonumber &&
+60uu_2^2+88u_1^2u_2+\frac{256}{5}u^3u_2+\frac{384}{5}u^2u_1^2
+\frac{1024}{125}u^5\bigg)
\end{eqnarray}


System (\ref{ex5}) is bi--Hamiltonian with Hamiltonian
operator:
\begin{eqnarray*}
{\cal H}_{11}&=&
\frac{7}{88}D_x^3+\frac{1}{55} (uD_x+D_x u)\\
\cH_{12}&=&\frac{11}{128}D_x^7+\frac{9}{20}D_x^3 u D_x^2
+\frac{11}{40} D_x^2 u_2D_x +\frac{6}{55}v D_x+\frac{2}{5} D_x^2 u^2D_x
+\frac{1}{11} u_t
\end{eqnarray*}
and a symplectic operator
\[\cJ_{21}=D_x^{-1}, \quad \cJ_{22}=0 \ . \]
Its restricted recursion operator is of order 6:
\begin{eqnarray*}
\Re&=&\frac{9}{128}D_x^6+\frac{21}{20}u D_x^4+\frac{12}{5}u_1 D_x^3
+\frac{17}{8}u_2 D_x^2+\frac{14}{5}u^2 D_x^2+\frac{9}{4}u_3 D_x
+\frac{72}{5}u u_1 D_x+\frac{51}{40} u_4\\
&&+8 u u_2 +\frac{42}{5} u_1^2 +\frac{6}{5} v+\frac{7}{8} D_x
D_t +\frac{2}{5} u D_x^{-1} D_t +\frac{1}{5} u_1 D_x^{-2} D_t+u_t
D_x^{-1}.
\end{eqnarray*}
Hierarchies of symmetries are of order $S_{6n+1}$ and $S_{6n+5}$
generated from the seeds $u_1$ and $u_t$ respectively.
There are no symmetries of even order and of order $6n+3, n\in \mathbb N$.

The Lax representation for the equation (\ref{ex5}) is given by:
\begin{eqnarray}
\nonumber
L&=&D^6+\frac{32}{5}uD^4+16u_1D^3+(\frac{96}{5}u_2+\frac{256}{25}u^2)D^2+(\frac{64}{5}u_3+\frac{768}{25}uu_1)D+
\frac{752}{225}u_4+\frac{256}{25}u_1^2+\\
\nonumber
&&+\frac{2816}{225}uu_2+\frac{8192}{3375}u^3+\frac{256}{225}v+(\frac{16}{225}u_5+\frac{256}{225}u_1u_2+
\frac{256}{225}uu_3+\frac{4096}{1125}u^2u_1+\frac{128}{225}v_1)D^{-1}\\
\nonumber
A&=&-\frac{9}{8}D^5-6uD^3-12u_1D^2-(10u_2+\frac{32}{5}u^2)D-4u_3-\frac{64}{5}uu_1\, .
\end{eqnarray}

After a simple invertible change of variables
\[
v=w-\frac{1}{8}u_4-2uu_2-\frac{32}{15}u^3,
\]
equation (\ref{ex5}) can be rewritten in the form
\begin{eqnarray*}
\left\{\begin{array}{l}
u_t=-\frac{1}{8}u_5-2uu_3-2u_1u_2-\frac{32}{5}u^2u_1+w_1\\
w_t=\frac{9}{8} w_5+6 u w_3+6 u_1 w_2+4 u_2 w_1 +\frac{32}{5} u^2 w_1
\end{array}\right. .
\end{eqnarray*}
It admits a reduction $w=0$ to the Sawada-Kotera equation \cite{sawadkotera}.

Much less is known about new integrable system (\ref{ex4b}). It can be rewritten in the form (\ref{eq1})
\begin{eqnarray}
\nonumber
u_{tt}&=&-\frac{1}{54}u_{10}+u_{5t}+D_x\left(\frac{5}{6}u_7u_1
+\frac{5}{3}u_6u_2+\frac{5}{2}u_5u_3+\frac{25}{12}u_4^2
-5u_3u_t-\frac{15}{2}u_2u_{1t}-10u_1u_{2t}\right) \label{ex4}\\
\nonumber&& +D_x\left(-\frac{45}{4}u_5u_1^2-\frac{75}{2}u_1u_2u_4
-\frac{75}{4}u_3^2u_1 -\frac{75}{4}u_2^2u_3+\frac{45}{2}u_1^2u_t+
\frac{225}{4}u_3u_1^3
+\frac{675}{8}u_2^2u_1^2-\frac{405}{16}u_1^5\right) \nonumber
\end{eqnarray}

System (\ref{ex4b}) is a bi--Hamiltonian system. We have found the following Hamiltonian operator
\begin{eqnarray*}
{\cal H}_{11}&=&
-\frac{14}{9}D_x+u_1D_x^{-1}+D_x^{-1} u_1\\
\cH_{12}&=&-\frac{41}{27}D_x^5+16u_1D_x^3+32u_2D_x^2+[\frac{89}{3}u_3-\frac{75}{2}u_1^2]D_x+\frac{41}{3}u_4-75u_1u_2+
5v_1D_x^{-1}+ D_x^{-1}v_1
\end{eqnarray*}
and a symplectic operator
\begin{eqnarray*}
\cJ_{12}&=&-\frac{1}{9}D_x^7+2u_1D_x^5+4u_2D_x^4+[u_3-\frac{15}{2}u_1^2]D_x^3-6u_4D_x^2+[-7u_5+30u_1u_3+\frac{45}{2}u_2^2+
6v_1]D_x-\\&&-3u_6+60u_2u_3+30u_1u_4-\frac{135}{2}u_1^2u_2+3v_2,\\
\cJ_{22}&=&-2D_x^3+3u_1D_x+3D_x u_1
\end{eqnarray*}

It follows from (\ref{recopHJ}) that the restricted recursion
operator for equation (\ref{ex4b}) constructed from the above
Hamiltonian and symplectic operators is of order $12$.
The hierarchies of symmetries can be generated
from the seed symmetries $S_{-1}=1$, $S_1=u_1$, $S_5=u_t$ and
$$
S_7=u_7-42 u_{2t}-42 u_1 u_5 -\frac{105}{2} u_3^2+189 u_1 u_t
+\frac{945}{2}u_1^2 u_3-\frac{2835}{8}u_1^4
$$
Equation (\ref{ex4b}) does not have symmetries of order $6n+3$ and even order
symmetries. We have not found the Lax representation
for this equation either.

System (\ref{ex4c}) does not represent a single non-evolutionary
equation (\ref{eq1}) in the standard set of dynamical variables,
since the right hand side of the second equation depends on the
variable $v$ explicitly. It can be written in the form (\ref{eq1})
after the extension of the set by a potential $z$, such that
$z_1=u$.

System (\ref{ex4c}) is bi-Hamiltonian. We have found the following Hamiltonian
\begin{eqnarray*}
&&\cH_{11}=-\frac{1}{2} D_x^3+ uD_x + D_x u \\
&&\cH_{12}=-\frac{1}{2}D_x^7+12 D_x^5 u -24 D_x^4 u_1+26 D_x^3 u_2
-80 D_x^3 u^2 -14 D_x^2 u_3+160 D_x^2 u u_1 +3 D_x u_4
\\&&
-60 D_x u u_2 -45 D_x u_1^2 +80 D_x u^3 +3 D_x v -v_1
\end{eqnarray*}
and a symplectic operators
\[\cJ_{21}=D_x^{-1}, \quad \cJ_{22}=0 \ . \]
The corresponding restricted recursion operator is of order $6$:
\[
\Re=\frac{1}{2}D_x D_t+u_2D_x^2-8 u^2 D_x^2-2 u D_x^{-1}D_t-16 u u_2
+3 u_1^2+112 u^3-3 v-2 u_t D_x^{-1}-6 u_1 D_x^{-1} (u_2-8 u^2)-u_1 D_x^{-2} D_t,
\]
which generates the higher symmetries from the seeds $u_1$ and $u_t$.

This equation has no symmetries of even order and of order $6n+3$. Its hierarchies of symmetries
can be generated from the seed symmetries $S_1=u_1$ and $S_5=u_t$.

After a simple invertible change of variables $v=w+u_4-20 uu_2-15 u_1^2+\frac{80}{3} u^3$
it can be rewritten in the form
\begin{eqnarray*}
\left\{\begin{array}{l}
u_t=u_5-20 uu_3-50 u_1u_2+80 u^2u_1+w_1\\
w_t=-6 u_3 w -2 u_2 w_1+96 u u_1 w+16 u^2 w_1
\end{array}\right. .
\end{eqnarray*}
Condition $w=0$ is an obvious reduction of this system. The reduced
equation is the well known Kaup-Kupershmidt equation
\cite{kaupkupershmidt}.

\section*{Summary}

Most of results obtained in the theory of integrable equations
concern the case of evolutionary equations. The theory of
nonevolutionary equations has new features. In order to study a
nonevolutionary equation we represent it by a system of evolutionary
equations. Such representation is not unique. A choice of the
representation determines a set of dynamical variables and the
corresponding differential ring. The concept of local symmetry,
which is central for the theory of integrable equations, can be
rigorously defined in the terms of this differential ring. Existence
or absence of infinite hierarchies of local symmetries may depend on
the choice of the representation, i.e. on the ring or its extension.
In contrast to the evolutionary case, the existence of one higher
symmetry does not guarantee the existence of an infinite hierarchy
of symmetries \cite{mnw}.

We have developed a perturbative symmetry approach suitable for
testing for integrability (testing necessary conditions for the
existence of higher local symmetries). It is based on the elements
of the standard symmetry approach and the symbolic representation of
the differential ring. Our approach proved to be very effective in
problems of testing and classification of integrable nonevolutionary
equations of orders 4,6 and 10. We believe that we have found a few
new integrable cases.  In all cases we have proven the integrability
by presenting a bi-Hamiltonian structure, a corresponding recursion
operator, a Lax representation or a suitable linearising
substitution. Our approach  can be developed even further, namely a
global, i.e. in all orders, classification of integrable equations
can be achieved (it would require some elements of the Number
Theory, such as the application of the Lech--Mahler Theorem and it is
beyond of the scope of this paper).

\section*{Acknowledgments}

We would like to thank  A.~Hone, E.L. Mansfield,
J.A.~Sanders and V.V.~Sokolov for useful discussions.
AVM is particularly grateful to the Vrije University,
Amsterdam, for Invited Visiting Thomas Stieltjes Professorial Fellowship,
which enable us to start this research project. AVM also thanks the RFBR for
a partial support (grant \# 05-01-00189). VSN was funded by a
Royal Society NATO/Chevening fellowship on the project {\it
Symmetries and coherent structures in non-evolutionary partial
differential equations} and is funded by the EU GIFT project (NEST --
Adventure Project no. 5006).

\end{document}